\documentclass[twocolumn]{article}
\usepackage[utf8]{inputenc}
\usepackage{abstract}
\usepackage{graphicx}
\usepackage{amsmath}
\usepackage{amssymb}
\usepackage{stfloats}
\usepackage{subcaption}
\usepackage{multirow}
\usepackage{bm}
\usepackage{wasysym}
\usepackage{xcolor,soul,colortbl}
\usepackage[most]{tcolorbox}

\definecolor{light-gray}{gray}{0.9}
\definecolor{myorange}{rgb}{1,0.48,0.19}
\definecolor{mysky}{rgb}{0,0.7,1}
\definecolor{mypink}{rgb}{.98,0.44,.84}
\definecolor{mylightgreen}{rgb}{0.85,1,0.7}
\definecolor{mylightred}{rgb}{1,0.8,0.8}
\definecolor{mygreen}{rgb}{0.53,.83,0.07}
\definecolor{myyellow}{rgb}{0.99,0.79,0.19}
\definecolor{mypurple}{rgb}{0.72,0.39,1}

\usepackage{hyperref}
\hypersetup{colorlinks,allcolors=mysky}

\usepackage[superscript]{cite}
\usepackage{geometry}
\usepackage{tabu}
\usepackage{makecell}
\usepackage{authblk}
\usepackage[font=small,labelfont=bf]{caption}
\usepackage{libertine}
\usepackage[T1]{fontenc}
\usepackage{inconsolata}   % Load the Inconsolata font

\tcbset{highlight math style={
  enhanced jigsaw,
  size=tight,
  colback=light-gray,
  boxrule=0pt,
  rounded corners,
  left=2pt,  
  right=2pt, 
  top=2pt,   
  bottom=2pt, 
  colframe=white,
  arc=3pt
}}

\newcommand{\code}[1]{\tcbhighmath{\small{\texttt{#1}}}}

\makeatletter  \renewcommand{\@biblabel}[1]{\textbf{#1}}  \makeatother

\title{New methods to compute the generalized chi-square distribution}

\author{Abhranil Das\\ \begin{footnotesize}\texttt{abhranil.das@utexas.edu}\end{footnotesize}}

\affil{Center for Perceptual Systems, and Center for Theoretical and Computational Neuroscience,\\
The University of Texas at Austin
}
\date{First published: Apr 7, 2024. Last revised: \today.}

\begin{document}
\maketitle

\begin{abstract}
We present four new mathematical methods, two exact and two approximate, along with open-source software, to compute the cdf, pdf and inverse cdf of the generalized chi-square distribution. Some methods are geared for speed, while others are designed to be accurate far into the tails, using which we can also measure large values of the discriminability index $d'$ between multivariate normal distributions. We compare the accuracy and speed of these and previous methods, characterize their advantages and limitations, and identify the best methods to use in different cases.

\end{abstract}
    
\section{Introduction}

The generalized chi-square variable $\tilde{\chi}$ is a quadratic function of a $d$-dimensional multivariate normal (henceforth called multinormal) variable $\bm{x} \sim N_d(\bm{\mu}, \bm{\Sigma})$, and can also be seen as a weighted sum of non-central chi-square variables $\chi'^2$ and a standard normal variable $z$:
\begin{align*}   \tilde{\chi}_{\bm{w}, \bm{k}, \bm{\lambda},s,m} &= q(\bm{x})=\bm{x}' \mathbf{Q}_2 \bm{x} + \bm{q}_1' \bm{x} + q_0\\
&= \sum_i w_i \, \chi'^2_{k_i,\lambda_i} + sz+m.
\end{align*}

$s$ is the scalar coefficient of the linear normal term, and $m$ is a constant offset. (Regular symbols are scalars, bold lowercase symbols are column vectors, bold uppercase symbols are matrices.)

This distribution arises across many fields, such as statistics and machine learning \cite{jones1983statistical,nishi2023counter,li2023adaptive,ahmed2023exact,haider2022unfair, wong2022prior, richmond2024estimation, manfredi2024probabilistic,cohen2024classifying, wang2024certified,haider2024identifying}, neuroscience  \cite{das2021method,das2020method}, cosmology \cite{hazboun2023analytic,allen2023hellings}, particle physics \cite{koch2025hypothesis}, signal transmission \cite{weiss2021detection, hsu2022statistical,graff2023purposeful, jin2024achievable,zhang2024interleaved,buonanno2024tolerance, tockner2024optimum}, satellite navigation \cite{rothmaier2021framework,kujur2024optimal}, radar \cite{bondre2024aspects,stockel2024optimized}, control theory \cite{balim2024stochastic,zhang2024numerical}, robotics \cite{frey2021belief,toner2022probabilistically,van2022provable,byeon2023stochastic}, target-tracking \cite{lewis2024improvements}, quality control \cite{manfredi2022probabilistic}, and cybersecurity \cite{liu2024linear}.

When $s=0$ and $w_i$ are all positive or negative, the quadratic is an ellipse. Then the distribution starts from the point $m$ at one end, which we call a finite tail. The other end tails off at + or $-\infty$ respectively, which we call an infinite tail. When $w_i$ have mixed signs, and/or there is a normal $s$ term, both tails are infinite.

There are several existing methods to compute the cdf and pdf of this distribution, which may behave differently in finite vs. infinite tails. When all $w_i$ are the same sign and $s=0$, i.e. the quadratic is an ellipse, Ruben's method \cite{ruben1962probability} can be used to compute the cdf, implemented as \code{gx2cdf($\ldots$, \textquotesingle method\textquotesingle,\textquotesingle ruben\textquotesingle)} in our Matlab toolbox \href{https://www.mathworks.com/matlabcentral/fileexchange/85028-generalized-chi-square-distribution}{`Generalized chi-square distribution'} (source code is \href{https://github.com/abhranildas/gx2}{on github}). We have also ported this to a \href{https://pypi.org/project/gx2/}{`gx2' python package} with the same functionality (source code is again on \href{https://github.com/abhranildas/gx2-py}{github}). Imhof \cite{imhof1961computing} used Gil-Pelaez's method \cite{gil1951note} of inverting the characteristic function to compute the cdf and pdf for mixed $w_i$ as well, but not with the normal term. Davies\cite{davies1973numerical} extended the characteristic function to let Imhof's inversion work with the normal $s$ term too, and we further incorporated the offset $m$ and implemented it as \code{gx2cdf($\ldots$, \textquotesingle method\textquotesingle,\textquotesingle imhof\textquotesingle)} and \code{gx2pdf($\ldots$, \textquotesingle method\textquotesingle,\textquotesingle imhof\textquotesingle)} in our toolbox. This method requires computing an integral, and we provide the option to compute it fast with double precision numerics, or slowly but more accurately using variable precision arithmetic (vpa).

These exact methods all work well in the center of the distribution, but far into the tails they begin to reach their limits of accuracy or speed at different points. For this reason several  approximations have been derived for the cdf, such as Imhof's extension \cite{imhof1961computing} of Pearson's approximation \cite{pearson1959note} (only usable when $s=0$), and Liu et al's \cite{liu2009new} and Zhang et al's \cite{zhang2022fast} approximations (only usable when $\bm{q_1}=0$, i.e. $s=0$, and $\mathbf{Q}_2$ is non-negative definite, i.e. when all $w_i$ are the same sign), which attempt to match the moments to those of some simpler distribution. These approximate methods are often only applicable to limited cases, and even then have their inaccuracies. See Zhang et al \cite{zhang2022fast}, Duchesne et al \cite{duchesne2010computing} and Bodenham et al \cite{bodenham2016comparison} for a review of these approximate methods and their limitations.

In this paper we develop four new methods, two exact and two approximate, to compute the generalized chi-square cdf and pdf. These methods will have different tradeoffs of speed and accuracy. Then we compare their performances on computing the cdf and pdf against existing exact methods. We show that together, they let us reach extremely small values in all tails of all types of generalized chi-square distributions, and we provide a table of the best method to use in each case. Next, we show further accuracy tests of the new methods by: (i.) replicating previously published cdf values, (ii.) by measuring the accuracy of our cdf and pdf methods against previous exact methods, and also the accuracy of our inverse cdf method, over a wide random sample of distributions, and (iii) by computing high values of the discriminability index $d'$ between two multinormals that can be checked by a simpler reference calculation.

\section{Mapping to a quadratic function} \label{sec:mapping}
We had previously  shown how to map from the parameters of a multinormal and the coefficients of its quadratic function to the parameters of the resulting generalized chi-square distribution  \cite{das2021method,das2020method} (function \code{norm\_quad\_to\_gx2\_params} in our toolbox). As a first step to some of our methods, it will help to find the inverse map, i.e. from the parameters of a generalized chi-square distribution to the parameters of a multinormal and its quadratic function. This inverse map is one-to-many: there are infinite pairs of multinormals and corresponding quadratics that have the same distribution. So, in order to arrive at a single solution, we shall consider a canonical form where the multinormal is the standard normal, and find a quadratic function of it that will produce the given generalized chi-square distribution.

Let us start from an example.  Suppose the generalized chi-square parameters are weights $\bm{w} = \begin{bmatrix} w_1 & w_2 & w_3 \end{bmatrix}$, degrees of freedom $\bm{k} = \begin{bmatrix} 1 & 1 & 2 \end{bmatrix}$, non-centralities $\bm{\lambda} = \begin{bmatrix} \lambda_1 & \lambda_2 & \lambda_3 \end{bmatrix}$, and linear coefficients $s$ and $m$. We build this as a sum of squares of independent scaled and shifted standard normal variables $z_i$, plus a linear term. For any term with multiple degrees of freedom, we split it into as many standard normals, choosing to put the entire non-centrality in the first term. So in this case we have the quadratic as:
\begin{multline}
   q(\bm{z})= w_1 (z_1-\sqrt{\lambda_1})^2 + w_2 (z_2 -\sqrt{\lambda_2})^2 \\
   + w_3 \{ (z_3 -\sqrt{\lambda_3})^2 + z_4^2\} + sz_5+m. \label{eq:quadform}
\end{multline}

Now let us express the quadratic in vector and matrix notation: $q(\bm{z})=\bm{z}' \mathbf{Q}_2 \bm{z} + \bm{q}_1' \bm{z} + q_0$, where $\bm{z}$ is a standard normal vector, $\mathbf{Q}_2$ is a square matrix, and $\bm{q}_1$ is a vector.
First consider $s=0$. Then the standard normal has $\sum k_i$ dimensions. Collecting the second order terms, we can see that the matrix $\mathbf{Q}_2$ is diagonal, here $\begin{bmatrix} w_1 & w_2 & w_3 & w_3 \end{bmatrix}$, i.e. in general it is constructed by appending each $w_i$ $k_i$ times. Then collecting the linear $z_i$ terms, we see that $\bm{q}_1' = \begin{bmatrix} -2 w_1 \sqrt{\lambda_1} & -2 w_2 \sqrt{\lambda_2} & -2 w_3 \sqrt{\lambda_3} & 0 \end{bmatrix}$, i.e. in general we append each $-2 w_i \sqrt{\lambda_i}$, followed by a 0 $k_i-1$ times. And $q_0 = \sum w_i \lambda_i +m$. Now when $s \neq 0$, we increase the dimension to $\sum k_i + 1$, append the diagonal $\mathbf{Q}_2$ with a 0, and append $\bm{q}_1$ with $s$. This mapping is available as function \code{gx2\_to\_norm\_quad\_params} in our toolbox.

Given any quadratic function of any multinormal, we can simplify it by first finding its generalized chi-square parameters, then mapping back to this canonical quadratic of the standard multinormal.

Even though in this inverse map we fix the multinormal to be the unit sphere, this is still not its only quadratic function that has the given distribution. What is the entire family of inverse maps? First, in a term with multiple degrees of freedom, we could split the total non-centrality in any way, which corresponds to rotating the quadratic around the origin in that sub-space. Second, we can apply any linear transform to the entire space where the multinormal and the quadratic lives. So we can spherically rotate the quadratic around the origin without changing the distribution, and finally we can scale and shift the space so that the multinormal is no longer standard.

By mapping from generalized chi-square parameters to quadratic coefficients, then sampling standard normal vectors and computing their quadratic function, we can sample from the generalized chi-square distribution (function \code{gx2rnd} in our toolbox). This is an alternative to sampling from the constituent noncentral chi-squares and a normal and adding them.

\begin{figure}[!t]
    \includegraphics[width=\columnwidth]{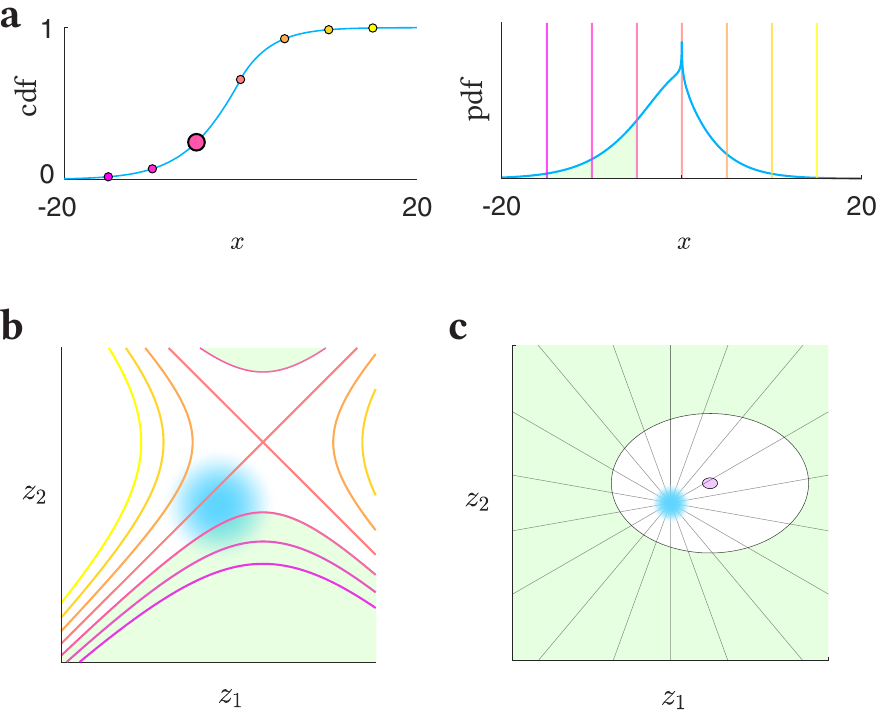}
    \caption{Mapping from the generalized chi-square parameters to the quadratic function of a multinormal, and integrating using ray-trace. \textbf{a.} A generalized chi-square cdf at one of several points (left, larger dot) is the pdf integrated upto that point (right, green area). \textbf{b.} This integrated probability is the standard multinormal (blue blob) probability over a domain (green area) that belongs to a family of quadratics (colours in this family correspond across plots a-b). \textbf{c.} The lower (finite) and upper (infinite) tail cdf's of a generalized chi-square that is a non-central ellipse, are the standard multinormal (blue blob) probabilities inside the tiny ellipse (purple area) and outside the large ellipse (green area).}
    \label{fig:1_mapping}
\end{figure}

Fig. \ref{fig:1_mapping}a shows the cdf of a generalized chi-square distribution with $\bm{w} = \begin{bmatrix} 1 & -1 \end{bmatrix}$, $\bm{k} = \begin{bmatrix} 1 & 1 \end{bmatrix}$, $\bm{\lambda} = \begin{bmatrix} 2 & 4 \end{bmatrix}$, $s=m=0$. To compute the cdf, we can first map these parameters to the corresponding quadratic function of a bivariate standard normal, $\tilde{\chi}=q(\bm{z})=z_1^2-z_2^2 - 2\sqrt{2}z_1 + 4z_2-2$, a hyperbolic function. The cdf at a point $F(c)=p(\tilde{\chi}<c)=p(q(\bm{z})<c)$ is then the multinormal probability within the contour of $q(\bm{z})$ at level $c$, i.e. the green area defined by the hyperbola $q(\bm{z}) < c$ or $q(\bm{z}) -c < 0$, fig. \ref{fig:1_mapping}b. The cdf at other points are the probabilities within the other hyperbolas in this family as we vary $c$. These probabilities can be integrated in the $\bm{z}$-space using the ray-tracing method, as described in the following section.

\section{Ray-tracing}

\subsection{Computing the cdf} \label{sec:ray-tracing}
Once we map from the generalized chi-square parameters to the quadratic function of a standard normal vector, we can now use the ray-tracing method \cite{das2021method,das2020method} to compute the cdf, available in our toolbox as \code{gx2cdf($\ldots$, \textquotesingle method\textquotesingle,\textquotesingle ray\textquotesingle)}. This is a method to compute the multinormal probability content over arbitrary domains, including quadratic domains, for which it is especially fast and accurate. In this method, we first linearly transform the space so that the distribution is the standard (unit spherical) multinormal. Then we send `rays' out from its center at every angle, and find the distances $z$ where each ray hits the integration domain, analogous to the ray-tracing method of rendering computer graphics. Using these intersection distances, we analytically integrate the density along each ray, which we then numerically integrate across all the rays. In Matlab we can use grid integration across rays for up to 4 dimensions, and Monte-Carlo integration above that (or even in general), which automatically runs in parallel on a GPU when present (here the NVIDIA GeForce RTX 3070). Ray-trace can use an onboard GPU only with double precision, and can be about 4 times faster than the CPU, but only when using at least about $10^6$ rays, due to overhead.

The ray method uses careful strategies to preserve precision and compute probabilities down to about $10^{-308}$, the smallest value expressible in double precision (called \code{realmin}). This is the case when its \code{precision} option is set to \code{basic}, and it already produces small enough probabilities for most imaginable applications. But for those beyond imagination, we now provide two more options to compute even smaller probabilities (these work only with Monte-Carlo integration across rays, not with Matlab's native grid integration).

The first of these is to set \code{precision} to \code{log}, which automatically approximates the log of probabilities on rays that are smaller than \code{realmin}, to avoid underflow. The probability on a ray, defined in our previous paper \cite{das2021method,das2020method}, involves computing the cdf $F_{\chi_d}(z)$ or complementary cdf $\bar{F}_{\chi_d}(z)$ of the chi distribution at the absolute distances $z$ where a ray from the origin intersects the integration domain. The tail probability, i.e. $F_{\chi_d}(z)$ at small $z$ or $\bar{F}_{\chi_d}(z)$ at large $z$, can fall below \code{realmin}, and there we write asymptotic expressions for their logs:
\begin{align*}
    \lim_{z \to 0} \log_{10} F_{\chi_d}(z) &= d \log_{10}z - \log_{10} \left[ d \ 2^{\frac{d}{2}-1} \ \Gamma \left(\frac{d}{2} \right)\right], \\
    \lim_{z \to \infty} \log_{10} \bar{F}_{\chi_d}(z) &= (d-2) \log_{10}z - \frac{z^2}{2 \ln 10} \\
    & - \log_{10} \left[ 2^{\frac{d}{2}-1} \ \Gamma \left(\frac{d}{2} \right)\right].   
\end{align*}

We can then sum these tiny probabilities across all the rays that contain them, using the `log-sum-exp' trick to sum in log-space, which starts from the logs of several small numbers, and computes the log of their sum, without underflow errors. This lets us compute the log of probabilities smaller than \code{realmin}, all the way down to the absurdly small value of $\log_{10} p = -\code{realmax} \approx -10^{308}$ (\code{realmax} is the largest double-precision number), i.e. $p = 10^{-10^{308}}$, all in fast double precision, and it can even use the GPU just like the \code{basic} option. We believe that nobody should ever need this level of precision, in the same way that Bill Gates allegedly said that nobody should need more than 640KB of computer memory.

The final option is to set \code{precision} to \code{vpa}. This implements the exact calculations (not approximations) symbolically, on the numbers that were too small for double precision, and evaluates the result with variable precision. With this setting, ray-tracing can be extended to compute probabilities smaller than \code{realmin}, all the way down to about $10^{-3 \times 10^{8}}$, where Matlab's symbolic calculation engine fails. Symbolic computation is slow and cannot use the GPU, so a quicker answer can only use fewer rays.

However, there is one situation in which the ray method cannot reach such small tail probabilities. When all $w_i$ are the same sign, $\sum \lambda_i > 0$ (i.e. the constituent chi-squares are not all central) and $s=0$, the corresponding quadratic function is an ellipse that is not at the origin. For example, consider $\tilde{\chi}=2\chi'^2_{k=1,\lambda=4}+4\chi'^2_{k=1,\lambda=1}$. This corresponds to the elliptical quadratic function: $\tilde{\chi}=q(\bm{z})=2(z_1-2)^2+4(z_2-1)^2$ of the bivariate standard normal (fig. \ref{fig:1_mapping}c). The lower (finite) tail probability $F(c)=p(q(\bm{z})<c)$, where $c$ is small, is the standard normal probability inside the tiny offset elliptical region (purple area), whereas the upper (infinite) tail probability $p(\tilde{\chi}>c)$, where $c$ is large, is the probability outside a large elliptical region (green area). When we send rays from the normal center to compute the upper tail probability, they all hit the large domain and return small values that accurately sum to the small upper tail probability. But in the lower tail, most rays miss the tiny elliptical domain. If the domain is very tiny, no ray may hit it, and the cdf will be incorrectly computed as 0. In $\leq 4$ dimensions, Matlab's adaptive grid integral can, to some extent, automatically find the tiny domain and populate the grid densely there and avoid this problem, but beyond 4 dimensions, randomly sampled Monte-Carlo rays will all miss a small enough domain. Therefore, the ray method is not the best for this situation, and we will develop the ellipse approximation that works well in this case. When the chi-square components are all central, the ray method does not have this problem. If there are many weights $w_i$, and nearly all of them are the same sign, and they are non-central, the domain is `nearly an offset ellipsoid', and still occupies a pretty small angle around the origin, so most rays will still miss it and this same problem will again occur there.

\subsection{Computing the pdf}

\begin{figure}[!t]
    \includegraphics[width=\columnwidth]{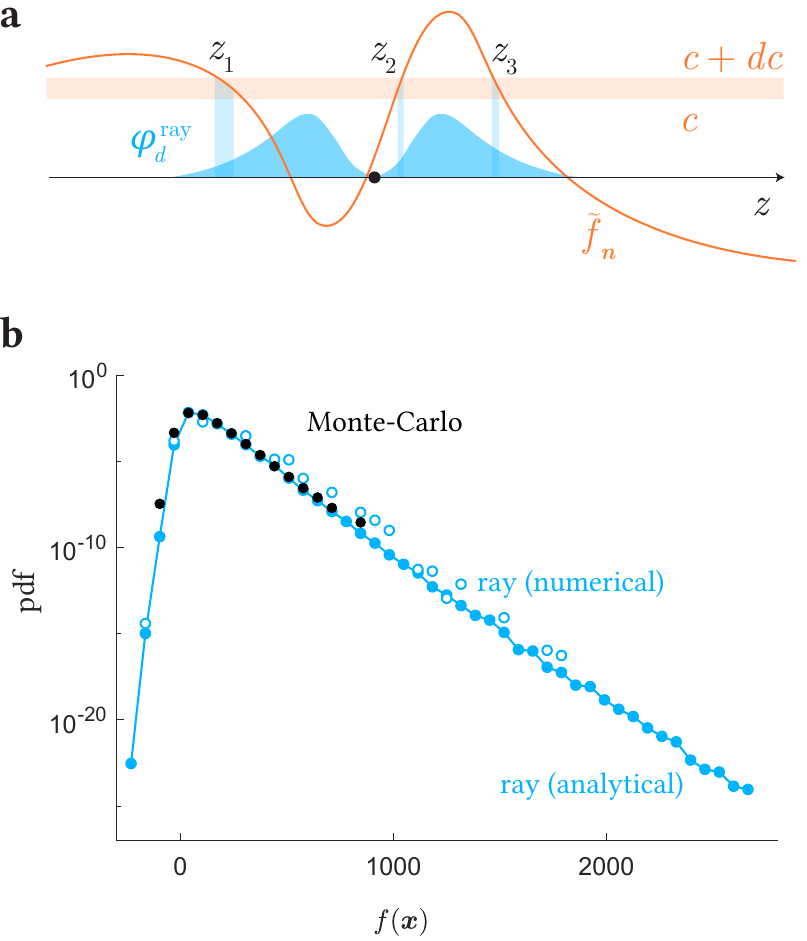}
    \caption{Ray-tracing method to compute the pdf of a general function $f(\bm{x})$ of a normal vector $\bm{x}$. \textbf{a.} The arrow is a ray through the mean (dot) of a standard multinormal along $\bm{n}$. Blue pdf $\phi^{\text{ray}}_d$ is the standard multinormal density along the ray. $\tilde{f}_{\bm{n}}(z)$ is the value of the standardized function $\tilde{f}(\bm{z})$ along the ray. Intervals where $\tilde{f}_{\bm{n}}$ lies between $c$ and $c+dc$ are the blue widths at $z_1$, $z_2$ and $z_3$. \textbf{b.} The pdf of a cubic function of a 4-dimensional correlated normal vector, computed by three methods in the same computation time. Missing dots are where a method wrongly computes the pdf as 0.}
    \label{fig:2_ray_pdf}
\end{figure}

The ray-tracing method, which we developed to compute the cdf of any function $f(\bm{x})$ of a normal vector $\bm{x} \sim N_d(\bm{\mu}, \bm{\Sigma})$, can be modified to compute the pdf too. A simple way to compute the pdf at $c$ is to compute the cdf at two nearby points $c \pm \Delta c$, and finite-difference them numerically. However, numerical differentiation is slower because it requires two evaluations of the cdf, and the differentiation itself introduces errors. Here we present the calculations to do this differentiation analytically instead, which extends the ray-tracing method to be able to compute densities of arbitrary functions of normal random vectors with greater speed and accuracy.

In the ray method, we first linearly transform the space so that the multinormal becomes the standard normal $\bm{z}$, and the function is transformed to the \textit{standardized} function in this space: $\tilde{f}(\bm{z})=f(\mathbf{S} \bm{z}+\bm{\mu})$, where $\mathbf{S}=\bm{\Sigma}^\frac{1}{2}$ is the symmetric square root. Similar to fig. 1 of our ray-tracing paper \cite{das2021method,das2020method} and corresponding explanations, fig. \ref{fig:2_ray_pdf}a here illustrates a ray sent from the origin (black dot) of the $d$-dimensional standard multinormal in the direction of a unit vector $\bm{n}$. $z$ is the coordinate along the ray, so that the vector location of point $z$ on the ray is $\bm{z}=\bm{n}z$ in the full space. The blue distribution $\phi^{\text{ray}}_d(z) = f_{\chi_d}(\lvert z \rvert)/2$ ($f_{\chi_d}$ is the chi distribution pdf) is the density of the standard multinormal along the ray, and $\tilde{f}_{\bm{n}}(z)$ is the value of the function $\tilde{f}$ along the ray.

Now, to find the pdf of $f(\bm{x})$ at $c$, we need to know the probability that $f(\bm{x})$, or $\tilde{f}(\bm{z})$ in the standardized space, is between $c$ and $c+dc$. This is the probability that the $\tilde{f}_{\bm{n}}(z)$ along each ray $\bm{n}$ is between $c$ and $c+dc$, summed across all rays. Fig. \ref{fig:2_ray_pdf}a shows this horizontal slice. The probability that $\tilde{f}_{\bm{n}}(z)$ is within this slice is the probability of finding $z$ within the corresponding vertical slices at $z_1$, $z_2$ or $z_3$ (the roots of $\tilde{f}_{\bm{n}}(z)-c$), given by $dp(\bm{n}) = \sum_i \phi^{\text{ray}}_d(z_i) \, \lvert dz_i \rvert $, where $\lvert d z_i \rvert$ is the width of each vertical slice, which can be found from the slope of the function: $\lvert \tilde{f}'_{\bm{n}}(z_i) \rvert = dc \big/ \lvert dz_i \rvert \implies \lvert d z_i \rvert = dc \big/ \lvert \tilde{f}'_{\bm{n}}(z_i) \rvert$. So we have $dp(\bm{n}) = \sum_i  \phi^{\text{ray}}_d(z_i) \big/ \lvert \tilde{f}'_{\bm{n}}(z_i) \rvert \, dc$. Now to integrate this across directions $\bm{n}$, we need to weight each $dp(\bm{n})$ by the volume fraction of the double-cone subtended by the differential angle element $d\bm{n}$, which is $\frac{2 d\bm{n}}{\Omega_d}$ ($\Omega_d$ is the total angle in $d$ dimensions), then sum. Since this gives us the probability that $f$ is between $c$ and $c+dc$, dividing away $dc$ then gives us the probability density:
\begin{equation*}
    \frac{2}{\Omega_d} \int_{\Omega_d/2} \underbrace{\sum_i  \frac{\phi^{\text{ray}}_d(z_i)}{\lvert \tilde{f}'_{\bm{n}}(z_i) \rvert}}_{\alpha(\bm{n})} \  d\bm{n}
\end{equation*}

For up to four dimensions, the integral is carried out numerically across half the total angle (since the probability on each ray already accounts for both directions of the ray). Beyond that, we perform a Monte Carlo integration, which finds the average value $\overline{\alpha}$ of the integrand over the \textit{total} angle $\Omega_d$, which is $\alpha(\bm{n})$ integrated over $\Omega_d$, divided by $\Omega_d$. Since the integral of $\alpha(\bm{n})$ over $\Omega_d$ is double that over $\Omega_d/2$, this automatically provides the extra factor of 2 needed to match the above equation.

Note that instead of inefficiently numerically computing the slope $\tilde{f}'_{\bm{n}}(z)$, we can see that it is simply the gradient of $\tilde{f}(\bm{z})$ in the direction $\bm{n}$: $\tilde{f}'_{\bm{n}} = \bm{n.\nabla} \! \tilde{f}$. This still needs the gradient of the standardized function, but we can at most expect the user to supply the gradient of the original function $f$. Fortunately, we can relate $\bm{\nabla} \! \tilde{f}$ to $\bm{\nabla}  \! f$. Remembering that $\tilde{f}(\bm{z})=f(\bm{x})$, where $\bm{x} = \mathbf{S} \bm{z}+\bm{\mu}$, i.e. $x_j=\sum_i S_{ji}z_i +\mu_j$, we can write:
\begin{align*}
    \nabla_i \tilde{f} (\bm{z}) &= \frac{\partial \tilde{f} (\bm{z})}{\partial z_i} = \frac{\partial f(\bm{x})}{\partial z_i} = \sum_j \frac{\partial f (\bm{x})}{\partial x_j} \frac{\partial x_j}{\partial z_i} \\
    &= \sum_j \nabla_j f (\bm{x}) \; S_{ji} = \sum_j S'_{ij} \nabla_j f (\bm{x}).
\end{align*}

That is, $\bm{\nabla} \! \tilde{f} (\bm{z}) = \mathbf{S}' \, \bm{\nabla} \! f(\mathbf{S} \bm{z}+\bm{\mu})$. So, when $f$ and $\bm{\nabla} \! f$ are both supplied, the pdf can be computed more quickly and accurately. This method is available as function \code{norm\_fun\_pdf} in our toolbox \href{https://www.mathworks.com/matlabcentral/fileexchange/84973-integrate-and-classify-normal-distributions}{`Integrate and classify normal distributions'} (source code is \href{https://github.com/abhranildas/IntClassNorm}{on github}).

Fig. \ref{fig:2_ray_pdf}b shows the pdf of a cubic function $f(\bm{x})=x_1^3+x_2^2-x_3 x_4$ of a 4-dimensional normal:
\begin{equation*}
    \bm{x} \sim N(\bm{\mu},\mathbf{\Sigma}), \quad \bm{\mu}= \begin{bmatrix} 4\\ -2\\ 3\\ 2
\end{bmatrix},
\mathbf{\Sigma} = \begin{bmatrix}
1 & 0 & -1 & 0\\
0 & 8 & 4 & 0\\
-1 & 4 & 8 & 0\\
0 & 0 & 0 & 1\\
\end{bmatrix},
\end{equation*}

computed using three methods, using settings such that they take the same time. There exists no other standard specialized method for computing pdf's of general functions of multinormals, so our baseline reference is a Monte-Carlo method where we sample normal vectors, compute their function values, then histogram them to estimate the pdf. With 5 million samples, this takes 0.2s per point on an 8-core 3.9GHz Intel Xeon W-2245 CPU (all CPU computation times reported henceforth are on this), and reaches down to a value of only about $10^{-9}$. The ray-tracing method with numerical differencing uses only 250 Monte-Carlo rays, and in the same time reaches down to about $10^{-17}$, but is noisy due to numerical errors in the differencing. The analytical derivative method uses 1000 rays in the same time and reaches down to about $10^{-25}$ with greater precision. The remaining small noisy errors in the pdf are due to the Monte Carlo sampling of rays, and can be smoothed by growing the sample. Note the distinction from the vanilla Monte-Carlo method: in ${>}4$ dimensions the ray method uses Monte-Carlo integration, but only across rays at different angles (the radial integral is computed analytically), so it converges much faster than using vanilla Monte-Carlo integration over all dimensions, as in the baseline method here.

Suppose we are using this method to find the pdf of a generalized chi-square distribution at the point $c$. Then we first find the corresponding quadratic $q(\bm{z})=\bm{z}' \mathbf{Q}_2 \bm{z} + \bm{q}_1' \bm{z} + q_0$ of the standard (multi)normal. Then for any ray in this space in the direction $\bm{n}$, the value of this function is a quadratic of the $z$ coordinate along the ray \cite{das2021method,das2020method}:
\begin{align*}
    q_{\bm{n}}(z)=q(z\bm{n})&=\bm{n}' \mathbf{Q}_2 \bm{n} z^2 +\bm{q}_1' \bm{n}z + q_0 \\
    &=q_2(\bm{n}) \, z^2 + q_1(\bm{n}) \, z + q_0.
\end{align*}

Now we find the roots of $q_{\bm{n}}(z)-c = q_2 z^2 + q_1 z + q_0 -c$. For there to exist two roots, we must have $q_2 \neq 0$ and the  discriminant $\Delta=q_1^2-4 q_2 (q_0-c) > 0$. Along the rays where this holds, the roots are $z_i = \frac{-q_1 \pm \sqrt{\Delta}}{2 q_2}$, and the slope of the quadratic is the same magnitude at either root: $\lvert q'_{\bm{n}}(z_i) \rvert = \lvert 2 q_2 z_i + q_1 \rvert = \sqrt{\Delta}$. And if $q_2 = 0$ but $q_1 \neq 0$, then the function is a line with one root $z=\frac{c-q_0}{q_1}$, with slope magnitude $\lvert q_1 \rvert = \sqrt{\Delta}$ again. These give us the quantities which we then integrate across different rays, either by quadrature (upto 4D) or Monte-Carlo, to get the generalized chi-square pdf at $c$. In our Matlab implementation \code{gx2pdf($\ldots$, \textquotesingle method\textquotesingle,\textquotesingle ray\textquotesingle)}, we implement this using fast vector operations, and speed up the Monte-Carlo integration by using the same sample of rays to compute the pdf at multiple points.

When the output pdf falls below \code{realmin}, we use the same method as with the cdf to compute its log. For this, we compute the log of the contribution from each root on each ray: $\log_{10}\phi^{\text{ray}}_d(z_i)-\log_{10} \lvert q'_{\bm{n}}(z_i) \rvert$, where the first term can be written exactly this time using the chi distribution pdf, unlike the case of the cdf. We then sum these using the `log-sum-exp' trick. With this we can compute densities down to $f = 10^{-10^{308}}$ in fast double precision, and even use a GPU.

\section{Inverse Fourier transform}
\subsection{Computing the cdf}
Arguably the best existing general method to compute the generalized chi-square cdf $F(x)$ is Imhof's method \cite{imhof1961computing}, which uses the Gil-Pelaez theorem \cite{gil1951note} to write it as an integral involving the characteristic function $\phi(t)$ (function \code{gx2char} in our toolbox):
\begin{equation*}
    F(x) = \frac{1}{2} - \frac{1}{\pi}\int_0^\infty \frac{\operatorname{Im} [\phi(t) \; e^{-itx}]}{t}\;dt.
\end{equation*}
This integral is then carried out numerically to the requested tolerance, which sums the integrand over an adaptive grid to approximate the cdf at the single point $x$.
However, let us follow a different path instead to get it to the form of a discrete inverse Fourier transform, starting from the Gil-Pelaez theorem:
\begin{align*}
    F(x) &= \frac{1}{2} +\frac{1}{2\pi}\int_0^\infty \frac{\phi(-t) \; e^{itx} - \phi(t) \; e^{-itx}}{it}\;dt \\
    &= \frac{1}{2} +\frac{1}{4\pi}\int_{-\infty}^\infty \frac{\phi(-t)\; e^{itx} - \phi(t) \; e^{-itx}}{it}\;dt \\
    &= \frac{1}{2} +\frac{1}{2\pi}\int_{-\infty}^\infty \frac{\phi(-t)}{it} \; e^{itx} \; dt.
\end{align*}
This gets it into the form of a continuous inverse Fourier transform of the function $\phi'(t) = \phi(-t) \big/ it$. Now, suppose $t_n= n \, \delta t, \ n = \{ -N, \dots, N \}$ is a uniform grid with spacing $\delta t$, from $-N \, \delta t$ to $N \, \delta t$. If $\delta t$ is small and $N \, \delta t$ is large, then we can approximate the integral as a discrete sum:
\begin{equation*}
    F(x) \approx \frac{1}{2} + \frac{1}{2\pi} \sum_{n=-N}^N \phi'(t_n) \; e^{i t_n x} \, \delta t.
\end{equation*}

Now, say we want to compute $F$ on a uniform grid of points centered at $x_\text{mid}$: $x_j= x_\text{mid} + j \, \delta x, \; j=\{ -N, \dots, N \}$, spanning a range $\Delta x = (2N+1) \, \delta x$.
Plugging in $x_j$ we get:
\begin{equation*}
    F(x_j) \approx \frac{1}{2} + \frac{1}{2\pi} \sum_{n=-N}^N \underbrace{\phi'(t_n) \; e^{i n \, \delta t \, x_\text{mid}}}_{\tilde{\phi}_n} \, e^{i n \, \delta t \, j \, \delta x} \, \delta t.
\end{equation*}
Now if we select $\delta t = \frac{2 \pi}{\Delta x} = \frac{2 \pi}{(2N+1) \, \delta x}$, this becomes:
\begin{equation*}
    F(x_j) \approx \frac{1}{2} + \frac{1}{(2N+1) \, \delta x} \sum_{n=-N}^N \tilde{\phi}_n \; e^{2 \pi i j n/(2N+1)} = \frac{1}{2} + \frac{\hat{\phi}_j}{\delta x},
\end{equation*}

where $\hat{\phi}_j$ is the discrete inverse Fourier transform of $\tilde{\phi}_n$. Most programming languages have fast optimized IFFT functions, using which we can compute the cdf at not one, but simultaneously the entire grid of points, which is an advantage over Imhof's method. However, a limitation of approximating a continuous Fourier transform with discrete is that since $\delta t$ has to be kept small, the span $\Delta x$ must be large to obtain accurate values of $F$ (in addition to a large number of grid points $N$), even when we only finally want them over a small range of $x$, and this sacrifices speed. There is another trade-off: the cdf is returned only over a uniform grid of points, whereas the Imhof-Davies method can compute it at any specific point. To do this here, we first evaluate it over a fine grid that surrounds those points, then interpolate to the given points.

This method is particularly good for a fast, moderately accurate computation of large parts of the distribution, say for quick plots. It is available as \code{gx2cdf($\ldots$, \textquotesingle method\textquotesingle,\textquotesingle ifft\textquotesingle)} in our toolbox.

\iffalse

Let's call $\tilde{\phi}_n= \phi(t_n)$. Now, suppose $x_j= j \Delta x, \; j=\{ -N, \dots, N \}$ is a uniform grid of points.
Substituting, we get:
\begin{equation*}
    F(x_j) \approx \frac{1}{2} + \frac{1}{2\pi} \sum_{n=-N}^N \tilde{\phi}_n \; e^{i n\Delta t \ j \Delta x} \, \Delta t.
\end{equation*}
So if we select $\Delta t = \frac{2 \pi}{(2N+1) \Delta x}$, this becomes:
\begin{equation*}
    F(x_j) \approx \frac{1}{2} + \frac{1}{\Delta x} \frac{1}{2N+1} \sum_{n=-N}^N \tilde{\phi}_n \; e^{2 \pi i j n/(2N+1)} = \frac{1}{2} + \frac{\hat{\phi}_j}{\Delta x},
\end{equation*}

\fi

\subsection{Computing the pdf}

The best existing method to compute the generalized chi-square pdf in general is also Imhof's method \cite{imhof1961computing}, which can be arrived at by differentiating the cdf expression, yielding:
\begin{equation*}
    f(x) = \frac{1}{\pi}\int_0^\infty \operatorname{Re} [\phi(t) \; e^{-itx}]\;dt.
\end{equation*}
Similar to the case of the cdf, we shall instead try to change this to a discrete inverse Fourier transform, starting from the basic inversion formula:
\begin{equation*}
    f(x)=\frac{1}{2\pi}\int_{-\infty}^\infty \phi(t)\; e^{-itx}\;dt = \frac{1}{2\pi}\int_{-\infty}^\infty \phi(-t)\; e^{itx}\;dt.
\end{equation*}
This is a continuous inverse Fourier transform of the function $\phi'(t)=\phi(-t)$. Following our previous derivation, we can approximate this as $f(x_j) \approx \hat{\phi}_j/\delta x$, where $\hat{\phi}_j$ is the discrete inverse Fourier transform of $\tilde{\phi}_n = \phi'(t_n) \; e^{i n \, \delta t \, x_\text{mid}}$.

This method is available as \code{gx2pdf($\ldots$, \textquotesingle method\textquotesingle,\textquotesingle ifft\textquotesingle)} in our toolbox.

\section{Extended Pearson approximation}
Imhof \cite{imhof1961computing} extended Pearson's approach \cite{pearson1959note} to approximate the generalized chi-square cdf with that of a central chi-square that matches its first three moments. However, this did not incorporate $s$ and $m$, which we will do here.

The first three moments of the generalized chi-square to be matched, that incorporate $s$ and $m$, are:
\begin{align*}
m_1 &= \sum_j w_j (k_j+\lambda_j) +m \quad \text{ (mean)}, \\
m_2 &= 2 \sum_j w_j^2 (k_j+2\lambda_j) + s^2  \quad \text{ (variance)}, \\
m_3 &= 8 \sum_j w_j^3 (k_j+3\lambda_j)  \quad \text{ (un-normalized skewness)}.
\end{align*}

We pick the degrees of freedom $k$ of the chi-square to match the skewness, which yields $k=8m_2^3/m_3^2$. Then we scale and shift the chi-square to match the mean and variance, which yields:
\begin{equation*}
    F(x) \approx F_{\chi^2_k} \left( (x-m_1)\sqrt{\frac{2k}{m_2}}+k \right),
\end{equation*}

from which we can also obtain an approximation for the pdf:
\begin{equation*}
    f(x) \approx \sqrt{\frac{2k}{m_2}} f_{\chi^2_k} \left( (x-m_1)\sqrt{\frac{2k}{m_2}}+k \right).
\end{equation*}

If $m_3 <0$, we flip the distributions, i.e. reverse the signs of $m_1$, $m_3$ and $x$, then use the same results above.

These methods are available in our toolbox as the functions \code{gx2cdf($\ldots$, \textquotesingle method\textquotesingle,\textquotesingle pearson\textquotesingle)} and \code{gx2pdf($\ldots$, \textquotesingle method\textquotesingle,\textquotesingle pearson\textquotesingle)}.

\section{Infinite-tail approximation}
\label{sec:inf-tail}

In this section we derive closed-form asymptotic expressions for the pdf and cdf in the infinite tails of the distribution.

The generalized chi-square pdf can be written as the Fourier inversion of its characteristic function $\phi(t)$:
\begin{equation*}
    f(x) = \frac{1}{2 \pi}\int_{-\infty}^{\infty} e^{-itx} \underbrace{e^{itm - s^2 t^2/2} \; \prod_j \frac{\exp \left(\frac{it w_j \lambda_j}{1-2iw_j t}\right)} {\left(1-2iw_jt \right)^{k_j/2}}}_{\phi(t)} \, dt
\end{equation*}

\begin{figure}[!t]
    \includegraphics[width=\columnwidth]{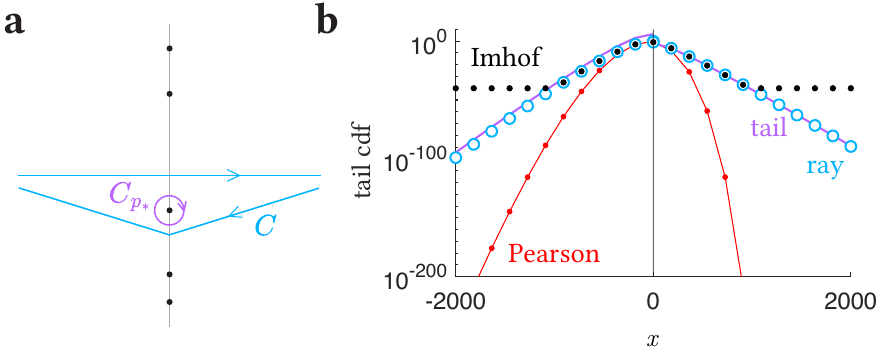}
    \caption{The infinite-tail approximation. \textbf{a.} Contour-integral derivation of the pdf. Black dots are singularities of the characteristic function. \textbf{b.} Comparing the infinite-tail (called simply `tail') method with several other methods in the far tails of the distribution.}
    \label{fig:3_inf_tail}
\end{figure}

The weights $w_j$ can have mixed signs, so the integrand has singularities $p_j=-i/2w_j$, both above and below 0, see fig. \ref{fig:3_inf_tail}a. Call the largest positive weight $w_*$, and its corresponding degree and non-centrality as $k_*$ and $\lambda_*$. Then the singularity in the lower half-plane that is nearest to the real line is $p_*=-i/2w_*$. We consider the case where $k_*$ is even.

This integral is rightward along the real line. Consider calculating this integral instead over a contour $C$ that goes rightward along the real line, then closes in the lower half plane below $p_*$, excluding all other singularities. We first show that as $x \rightarrow \infty$, this contour integral approximates the desired integral over just the real line. Suppose we are in the lower-half-plane, so that $t=u-iv$, $v>0$. The integrand is then $e^{-vx} e^{-iux} \phi(t)$. So as $x \rightarrow \infty$, at any point in the lower-half-plane where $\phi(t)$ stays finite (i.e. excluding points infinitely far down, and singularities of $\phi(t)$), the integrand becomes infinitesimal relative to its value along the real line where $v=0$. So the contribution of the lower part of the contour $C$ vanishes relative to the horizontal integral that we want.

The integral over $C$ equals that along a small clockwise circle $C_{p_*}$ around $p_*$. We re-write the integral as:
\begin{equation*}
    \oint_{C_{p_*}} e^{-itx} \underbrace{e^{itm - s^2 t^2/2} \; \prod_{j \neq *} \frac{\exp \left(\frac{it w_j \lambda_j}{1-2iw_j t}\right)} {\left(1-2iw_jt \right)^{k_j/2}}}_{a(t)} \; \frac{\exp \left(\frac{it w_* \lambda_*}{1-2iw_* t}\right)} {\left(1-2iw_*t \right)^{k_*/2}} \, dt
\end{equation*}

In the limit of the circle becoming infinitesimal, $a(t) \rightarrow a(p_*)$, which is a constant:
\begin{equation*}
    a(p_*) \oint_{C_{p_*}} e^{-itx} \frac{\exp \left(\frac{it w_* \lambda_*}{1-2iw_* t}\right)} {\left(1-2iw_*t \right)^{k_*/2}} \, dt.
\end{equation*}

Integrating this function along $C_{p_*}$ equals integrating it along a semicircular contour of radius $R$ that goes right along the real line and closes in the lower half plane. As $R \rightarrow \infty$, the integrand on the arc vanishes, leaving the integral over the real line. With $t'=w_*t$, this becomes:
\begin{equation*}
    \frac{a(p_*)}{w_*} \int_{-\infty}^{\infty} e^\frac{-it'x}{w_*} \frac{\exp \left(\frac{it' \lambda_*}{1-2it'}\right)} {\left(1-2it' \right)^{k_*/2}} \, dt'.
\end{equation*}
This is the Fourier inversion of the non-central chi-square characteristic function, evaluated at $x/w_*$. Plugging in $a(p_*)$ we get:
\begin{gather*}
 \lim_{x \rightarrow \infty} f(x) = \frac{a}{w_*} f_{\chi'^2_{k_*,\lambda_*}} \left(x/w_* \right), \\
 \text{where } a = e^{\frac{m}{2w_*} + \frac{s^2}{8w_*^2}} \prod_{j \neq *} \frac{\exp{\frac{\lambda_j w_j}{2(w_*-w_j)}}}{\left(1-\frac{w_j}{w_*} \right)^{k_j/2}}.
\end{gather*}

Note that as $x \rightarrow \infty$, the central or non-central chi-square pdf term here is dominated by an exponential decay $e^{-x/2w_*}$. Including the other lower-half-plane singularities with even order in our initial contour $C$ would add their residues, which are similar exponential decays, to our answer. But this term for the largest weight has the slowest decay and dominates, so we ignore the rest. If any of these other singularities have odd order, they are branch points and we cannot encircle them in an integration contour anyway.

From the pdf we can write the asymptotic cdf in the upper tail:
\begin{equation*}
 \lim_{x \rightarrow \infty} \bar{F}(x) = a \ \bar{F}_{\chi'^2_{k_*,\lambda_*}} \left(x/w_* \right).
\end{equation*}

When $\lambda_*=0$, the non-central chi-square pdf and cdf terms are replaced by the central ones.

Now to find the behaviour of $f(x)$ in the lower tail, we can close the contour in the upper half-plane and take the limit of $x \mathrel{\rightarrow} -\infty$ in all the calculations. Alternatively, we can simply flip the distribution horizontally (i.e. flip the signs of $w_j$ and $m$), and follow the same procedure above for the upper tail. This leads us to the same result for the pdf, except $w_*$ is now the largest negative weight, and the pre-factor is $-a/w_*$. For the lower tail cdf, we arrive at the same expression as before. So together we can say,
\begin{align*}
 \lim_{x \rightarrow \pm \infty} f(x) &= \frac{a}{\lvert w_* \rvert} f_{\chi'^2_{k_*,\lambda_*}} \left(x/w_* \right), \\
 \lim_{x \rightarrow -\infty} F(x) = \lim_{x \rightarrow \infty} \bar{F}(x) &= a \ \bar{F}_{\chi'^2_{k_*,\lambda_*}} \left(x/w_* \right) \\
 & = a \ Q_{k_*/2}(\sqrt{\lambda_*},\sqrt{x/w_*}),
\end{align*}

where $Q$ is the Marcum Q-function.

When $k_*$ is odd, $p_*$ is a branch point and the derivation cannot be done in the same way of a contour integral encircling  it. However, we have seen numerically that the same answer still holds. A similar case is where McNolty has deduced the non-central chi-square pdf by contour-integrating its characteristic function \cite{mcnolty1962contour}, having to use a more complicated derivation when the singularity is of odd order, but arriving at the same answer. The derivation here for odd $k_*$, which we leave as future work for those interested, could potentially follow such an approach, although there will be comparatively greater complications owing to the presence of the other singularities, and the fact that $\tilde\chi$ can be negative too.

We can compute the pdf and cdf expressions derived here even when they fall below \code{realmin}, by evaluating their logs. First, for the pdf, when $\lambda_*=0$, i.e. the approximation uses the central chi-square pdf, we can write:
\begin{multline*}
    \lim_{x \rightarrow \pm \infty} \log_{10} f(x) = \log_{10} \frac{a}{\lvert w_* \rvert} - \log_{10} \left[ 2^{\frac{k_*}{2}} \Gamma\left(\frac{k_*}{2} \right) \right] \\ + \frac{k_*-2}{2} \log_{10} \frac{x}{w_*} -\frac{x}{2w_* \ln 10}.
\end{multline*}

Now for $\lambda_*>0$, the noncentral chi square pdf involves a modified Bessel function of first kind, and we can use an asymptotic approximation for it, $\lim_{x \rightarrow  \infty} I_{\nu}(x)=e^x/\sqrt{2\pi x}$, to write:
\begin{multline*}
    \lim_{x \rightarrow \pm \infty} \log_{10} f(x) = \log_{10} \frac{a}{\lvert w_* \rvert} + \log_{10} \frac{\lambda_* ^{\frac{1-k_*}{4}}}{2 \sqrt{2 \pi}} \\
    + \frac{k_*-3}{4} \ \log_{10} \frac{x}{w_*} +\frac{1}{\ln 10} \left( \sqrt{\frac{\lambda_* x}{w_*}} -\frac{\frac{x}{w_*}+\lambda_*}{2} \right).
\end{multline*}

Now for the tail cdf's. When $\lambda_*=0$, the approximation uses the survival function of the central chi-square, which is an upper incomplete gamma function. We can use an asymptotic expression for this to write the $\log_{10}$ of the tail cdf's as:
\begin{equation*}
    \log_{10}a + \frac{k_*-2}{2} \ \log_{10} \frac{x}{2w_*} -\frac{x}{2w_* \ln 10} - \log_{10} \Gamma \left(\frac{k_*}{2} \right).
\end{equation*}

When $\lambda_*>0$, we can use an asymptotic approximation for the Marcum Q-function in terms of the standard normal survival function $\bar{\Phi}(z)$ \cite{simon2004digital}, which in turn can also be approximated at large $z$:
\begin{align*}
    \lim_{b \rightarrow  \infty} Q_m(a,b) &= \left(\frac{b}{a}\right)^{m-1/2} \  \bar{\Phi}(b-a), \\
    \lim_{z \rightarrow  \infty} \bar{\Phi}(z) &= \frac{1}{z \sqrt{2 \pi}} e^{-\frac{z^2}{2}}.
\end{align*}

Using these, we can write the $\log_{10}$ of the tail cdf's as:
\begin{equation*}
    \log_{10} \frac{a \lambda_*^{\frac{1-k_*}{4}}}{\sqrt{2 \pi}} + \frac{k_*-3}{4} \ \log_{10} \frac{x}{w_*} -\frac{\left(\sqrt{\frac{x}{w_*}}-\sqrt{\lambda_*}\right)^2}{2 \ln 10}.
\end{equation*}

These let us compute the tail pdf and cdf's all the way down to $10^{-10^{308}}$, in fast double precision, just like the \code{log} option of the ray method.

This method is available as \code{gx2cdf($\ldots$, \textquotesingle method\textquotesingle,\textquotesingle tail\textquotesingle)} and \code{gx2pdf($\ldots$, \textquotesingle method\textquotesingle,\textquotesingle tail\textquotesingle)} in our toolbox.

Looking at the logarithmic expressions, if we take only the terms involving $x$ (which dominate in the far tail), we can write even simpler asymptotic expressions which are then identical for the pdf $f$ and tail cdf $p$:
\[
f(x) \approx p(x) \approx 
\begin{cases}
\left(x/w_*\right)^{\frac{k_*-2}{2}} e^{-x/2w_*}, & \text{if } \lambda_*=0, \\[1ex]
\left(x/w_*\right)^{\frac{k_*-3}{4}} e^{-x/2w_* + \sqrt{\lambda_* x/w_*}}, & \text{if } \lambda_*>0.
\end{cases}
\]

These serve as extremely simple and fast rough asymptotic approximations for the very far tail probability and density of the generalized chi-square. Even within these simple expressions, only the exponential decays will further dominate in the far tail.

Fig. \ref{fig:3_inf_tail}b compares the infinite-tail approximation of the cdf with several other methods. Imhof's method with variable precision takes 28 sec to reach down to a value of $10^{-50}$, where it flatlines. Ray method takes the same time, but reaches much farther down. The tail approximation is consistent with these exact methods in the tails, and it takes only 16 ms.

Moment-matching methods, such as Pearson's \cite{pearson1959note,imhof1961computing} and Liu et al's \cite{liu2009new} approximations, also arrive at a central or non-central chi-square form like the tail approximation. However, in the far tails of fig. \ref{fig:3_inf_tail}b, we see that Pearson's method, for example, does not correctly match even the dominant exponential falloff rate, which ought to be the very first order approximation. This is because Pearson's method was designed by matching moments, which is good at matching the shapes of two distributions over finite spans, whereas the infinite-tail approximation is explicitly designed to match their asymptotic shapes at infinity.

\section{Finite-tail ellipse approximation}

\subsection{Computing the cdf}
As we mentioned in the introduction, the generalized chi-square distribution has a finite tail when $w_i$ are all the same sign and $s=0$, i.e. when the quadratic is an ellipse. For example, consider all positive $w_i$ and $s=m=0$. The lower tail at 0 here is such a finite tail. Imhof's method is not very good at computing small cdf values in this tail, and as we discussed, neither is the ray method unless the components are all central. Ruben's method\cite{ruben1962probability}, which applies specifically to this situation, can be used to compute the cdf as a sum of chi-square cdf's:
\begin{equation*}
    F(x)=\sum_i a_i F_{\chi^2_{d+2i}} (x/\beta).
\end{equation*}

This method performs better in this tail and can reach down to \code{realmin} pretty fast, below which it returns 0. For the upper tail (complementary cdf), if we subtract the cdf from 1, it leaves a numerical residue and never gets smaller than about $10^{-16}$. Instead, we can use a large number of terms in the Ruben series, and knowing that in this limit the $a_i$  should sum to 1, we can write:
\begin{equation*}
    \bar{F}(x)=\sum_i a_i \bar{F}_{\chi^2_{d+2i}} (x/\beta).
\end{equation*}

This lets us go down to \code{realmin} in the upper tail as well.

We can also differentiate Ruben's method to obtain an expression for the pdf in this case as a sum of chi-square pdf's:
\begin{equation*}
    f(x)=\frac{1}{\beta} \sum_{i=0}^{\infty} a_i f_{\chi^2_{d+2i}} (x/\beta).
\end{equation*}

These are available as \code{gx2cdf($\ldots$, \textquotesingle method\textquotesingle,\textquotesingle ruben\textquotesingle)} and \code{gx2pdf($\ldots$, \textquotesingle method\textquotesingle,\textquotesingle ruben\textquotesingle)} in our Matlab toolbox.

Now in this section we show a simple approximation that can also be used for such a finite tail, which can reach much lower than Ruben's method and also becomes exact in that limit.

Note that in this case the quadratic function corresponding to the distribution is an ellipse (or an ellipsoid or hyper-ellipsoid in general), that gets vanishingly small as we approach the end of the tail, e.g. the purple ellipse in fig. \ref{fig:1_mapping}c. The cdf is given by the probability inside this ellipse:

\begin{equation*}
    F (x) = p\left(\sum_{j=1}^d \omega_j (z_j-c_j)^2 < x \right).
\end{equation*}

The parameters of this ellipse can be easily identified once we write the quadratic function like in eq. \ref{eq:quadform}. The ellipse is in $d=\sum k_i$ dimensions, and its center $\bm{c}$ is the vector built by appending each $\sqrt{\lambda_i}$ followed by a 0 $k_i-1$ times. The weights $\omega_j$ are the elements of the diagonal matrix $\mathbf{Q}_2$ derived in sec. \ref{sec:mapping}, formed by appending each $w_i$ $k_i$ times. The semi-axis-lengths are $a_j=\sqrt{x/\omega_j}$.

A simple approximation for the probability inside this ellipse is to first compute the probability in the (hyper-) rectangle that encloses it (whose vertices are semi-axis-lengths above and below the center), using, say, Matlab's \code{mvncdf}, then multiplying it by the ratio of the ellipse volume to the rectangle volume. The approximation here is that we take the average probability density over the ellipse to be the same as over the rectangle, which becomes increasingly accurate as we approach the tail end and the ellipse gets tinier. However, computing the probability in the rectangle has its own associated error, and it cannot even always reach \code{realmin} like Ruben's fast exact method can, so this approximation is not very useful.

An even simpler approximation is to multiply the multinormal probability density at the ellipse center $\bm{c}$, with the ellipse volume:
\begin{align} \label{eq:4_ellipse_cdf}
    \lim_{x \to 0} F (x) &= (2 \pi)^{-\frac{d}{2}} \ e^{-\frac{\lVert \bm{c} \rVert^2}{2}} \ \frac{\pi^{\frac{d}{2}} \sqrt{\prod_j \frac{x}{\omega_j}}}{\Gamma \left(\frac{d}{2}+1 \right) } \nonumber \\
    &= \frac{e^{-\frac{\lVert \bm{c} \rVert^2}{2}}}{\Gamma \left( \frac{d}{2}+1 \right) \sqrt{\prod_j \omega_j}} \; (x/2)^{d/2}.
\end{align}

So the cdf is simply proportional to a power of $x$. Here the approximation is that we take the average probability density over the ellipse to be the same as at its center, which again becomes more accurate as we approach the tail end.

This expression corresponds to  the first term ($j=k=0$) of the power series expression of $F(x)$ that Shah and Khatri \cite{shah1961distribution} derived for this case, albeit not from a geometric interpretation such as this.

To better compute tiny cdf values at tiny $x$ values, we can look at the log probability instead, which is simply linear in log $x$:
\begin{multline*}
    \lim_{x \to 0} \log_{10} F(x) = \frac{d}{2} \left( \log_{10} x - \log_{10} 2 \right) - \frac{\lVert \bm{c} \rVert^2}{\ln 100} \\
    - \log_{10} \left[\Gamma \left (\frac{d}{2}+1 \right) \right] - \frac{1}{2} \sum_j \log_{10} \omega_j.
\end{multline*}

This lets us directly compute the log of tiny tail probabilities given the log of tiny $x$ values, all the way down to $F(x) = 10^{-10^{308}}$, in fast double precision.

We can also set lower and upper error bounds $F_{\text{min}}$ and $F_{\text{max}}$ for this cdf estimate, as the ellipse volume times the lowest and highest multinormal densities on the ellipse, i.e. at the points $\bm{e}_f$ and $\bm{e}_n$ on the ellipse that are farthest and nearest to the origin respectively. (The estimate is closer to the true value than this bound indicates, and even tighter bounds could be found, but this one is good enough as a first simple calculation.) When $\sum \lambda_i >0$, i.e. the components are not all central, the ellipse is not at the origin, and these two points are where the line from the origin through the ellipse center intersects the ellipse. So they are equidistant from the ellipse center on opposite sides of it, of the form $\bm{c} (1 \pm r)$, and $r$ is found by requiring that they satisfy the equation of the ellipse:
\begin{equation*}
    \bm{e}_f, \bm{e}_n = \bm{c} (1 \pm r), \quad r=\sqrt{\frac{x}{\sum_j c_j^2 \omega_j}}.
\end{equation*}

Using these points we can express the bounds as fractions of the estimate itself:
\begin{align*}
     \frac{F_{\text{min}}}{F} &= e^{\left(\lVert \bm{c} \rVert^2 - \lVert \bm{e}_f \rVert^2\right)/2} = e^{-\lVert \bm{c} \rVert^2 \left(r^2+2r \right)/2} \\
     \implies \frac{\Delta F^{-}}{F} &= \frac{F - F_{\text{min}}}{F} = 1-e^{-\lVert \bm{c} \rVert^2 \left(r^2+2r \right)/2}, \\
     \text{similarly } \frac{\Delta F^{+}}{F} &= \frac{F^{\text{max}}-F}{F} = e^{-\lVert \bm{c} \rVert^2 \left(r^2-2r \right)/2}-1.
\end{align*}

If these relative errors were constants, as we approach the tail end and the estimate goes to 0, the relative error would stay the same fraction of the estimate. But as $x \to 0$, $r \to 0$, and these estimation errors, even as a fraction of the estimate, go to 0 and the estimate becomes exact.

We can simplify the expressions for the relative errors at the limit, using the Maclaurin series of $r$ and taking only up to the linear term, which gives us symmetric error bounds:
\begin{equation*}
    \lim_{x \to 0} \frac{\Delta F^{-}}{F} = \lim_{x \to 0} \frac{\Delta F^{+}}{F} = \lVert \bm{c} \rVert^2 r = \lVert \bm{c} \rVert^2 \sqrt{\frac{x}{\sum_j c_j^2 \omega_j}}.
\end{equation*}

We can again take the log of this to be able to compute small relative error bounds using the log of tiny $x$ values.

Now, when each $\lambda_i=0$, i.e. all the components are central, the ellipse is at the origin, and the points on the ellipse that are nearest and farthest from the origin are respectively its center, i.e. the origin itself, and the point at the tip of its longest axis, at a distance $\sqrt{x/\omega_{\text{min}}}$.
\iffalse
Here the ellipse center has the maximum instead of an intermediate density, so for a better cdf estimate, we instead take the point $\bm{a}_{\text{max}}/\sqrt{2}$, and multiply its density with the ellipse volume. The choice of this point means that as $x \to 0$, the relative errors are symmetric around it: $\Delta F / F = x/4\omega_{\text{min}}$.
\fi
So if we use the density at the ellipse center for the cdf estimate, that is also its upper bound, and we can follow similar calculations as before to say that the lower relative error $\frac{\Delta F^{-}}{F} = 1-e^{-x/2\omega_{\text{min}}}$, and as $x \to 0$, this goes as $x/2\omega_{\text{min}}$, so now it is proportional to $x$ instead of $\sqrt{x}$ and is tighter.

We can also invert the formula for the error to decide where we can reliably use the ellipse approximation. For a non-central ellipse, if we want a relative error of no more than $\delta$, we can use the approximation in the range $x< \frac{\delta^2}{\lVert \bm{c} \rVert^4} \sum_j c_j^2 \omega_j.$ For example, take $\bm{w} = \begin{bmatrix} 3 & 1 & 2 \end{bmatrix}$, $\bm{k} = \begin{bmatrix} 4 & 2 & 3 \end{bmatrix}$, $\bm{\lambda} = \begin{bmatrix} 7 & 0 & 2 \end{bmatrix}$, $s=m=0$. For a relative error of ${<}1\%$, we can use the approximation for $x < 3 {\times} 10^{-5}$. For a central ellipse, we can use the approximation for $x < 2 \delta \omega_{\text{min}}$. For example, if all $\lambda_i=0$ in this example, we can use the approximation for $x < 0.02$.

\begin{figure}[t]
    \includegraphics[width=\columnwidth]{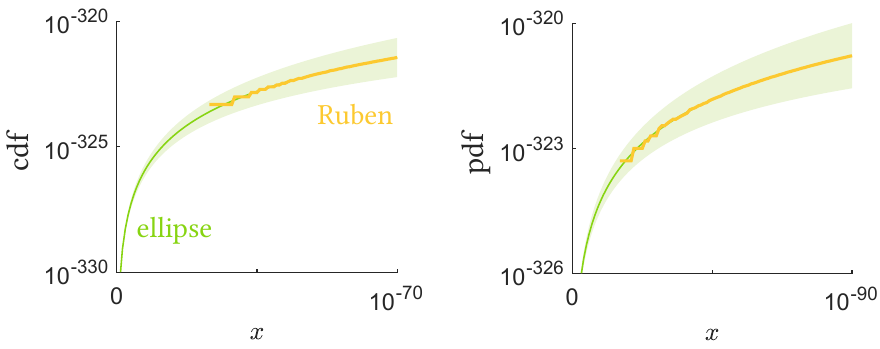}
    \caption{Comparing the ellipse approximation with Ruben's method for computing the cdf (left) and pdf (right) in the finite tail. Error-bands represent $10^{35}$ times the error for the ellipse cdf estimate, and $10^{45}$ times the error for the ellipse pdf estimate.}
    \label{fig:4_ellipse}
\end{figure}

Fig. \ref{fig:4_ellipse}, left, compares the ellipse cdf estimate to Ruben's method for the above non-central parameters, in the range where the computed cdf is around \code{realmin}, so at some point Ruben's method falls to 0 and disappears in this plot. We compute the log of the Ruben cdf for this plot, and log becomes an erroneous operation at such low values, so the Ruben curve zig-zags before disappearing. For the ellipse estimate we can directly compute the log, and continue much further down without any numerical issues. The error-band shows $10^{35}$ times the error around the ellipse estimate, so the error is actually miniscule. The way we plotted these error-bands of $F \pm \alpha \Delta F$ ($\alpha =10^{35}$) on a log axis is the following. Recall that the ellipse method returns the relative estimation error $\delta$. Then we can write: 
\begin{equation*}
    \log_{10} (F \pm \alpha \Delta F) = \log_{10} F + \log_{10} (1 \pm \alpha \delta) \approx \log_{10} F \pm \frac{\alpha \delta}{\ln 10},
\end{equation*}

where in the second step we assume small $\delta$, and have taken up to the linear term of its Maclaurin series.

In summary, we can see that the ellipse method agrees very well with Ruben's method in this tail, and helps to extend it down to even smaller cdf values with tightening accuracy.

These results can be easily extended to the case of all-negative $w_i$, where the distribution is flipped and the upper tail is finite, and also when the offset is shifted to a non-zero $m$. This method is available as \code{gx2cdf($\ldots$, \textquotesingle method\textquotesingle,\textquotesingle ellipse\textquotesingle)} in our toolbox.

\subsection{Computing the pdf}
The ellipse approximation can also provide an expression for the pdf in the finite tail, by simply differentiating the cdf, eq. \ref{eq:4_ellipse_cdf}, w.r.t. $x$:
\begin{equation} 
    \lim_{x \to 0} f(x) = \frac{d e^{-\frac{\lVert \bm{c} \rVert^2}{2}}}{2^{\frac{d}{2}+1}\Gamma \left( \frac{d}{2}+1 \right) \sqrt{\prod_j \omega_j}} \; x^{\frac{d}{2}-1}.
\end{equation}

We can again take the log of this to be able to compute the log of tiny density values at tiny $x$ values, down to $f(x) = 10^{-10^{308}}$. We can also set error bounds here again. The pdf approximation can be thought of as the rate of change of the ellipse volume with $x$, multiplied with the density at the ellipse center. As before, to get the error bounds we can instead multiply with the densities at the ellipse points closest and farthest from the origin. This leads to the same relative errors as for the cdf, for both the non-central and central cases.

This method is available as \code{gx2pdf($\ldots$, \textquotesingle method\textquotesingle,\textquotesingle ellipse\textquotesingle)} in our toolbox.

Fig. \ref{fig:4_ellipse}, right, compares the pdf computed by the ellipse method and Ruben's method (which is the best pdf method for the finite tail), for the same parameters again. The error-band here shows $10^{45}$ times the error of the ellipse method, so it is again tiny. So for the pdf too, the ellipse method agrees well with Ruben's method and extends it much farther into the tail.

\section{Comparing the methods}

\subsection{Computing the cdf}

\begin{figure*}[ht]
    \centering
    \begin{subfigure}[b]{\textwidth}
        \centering
        \includegraphics[width=\linewidth]{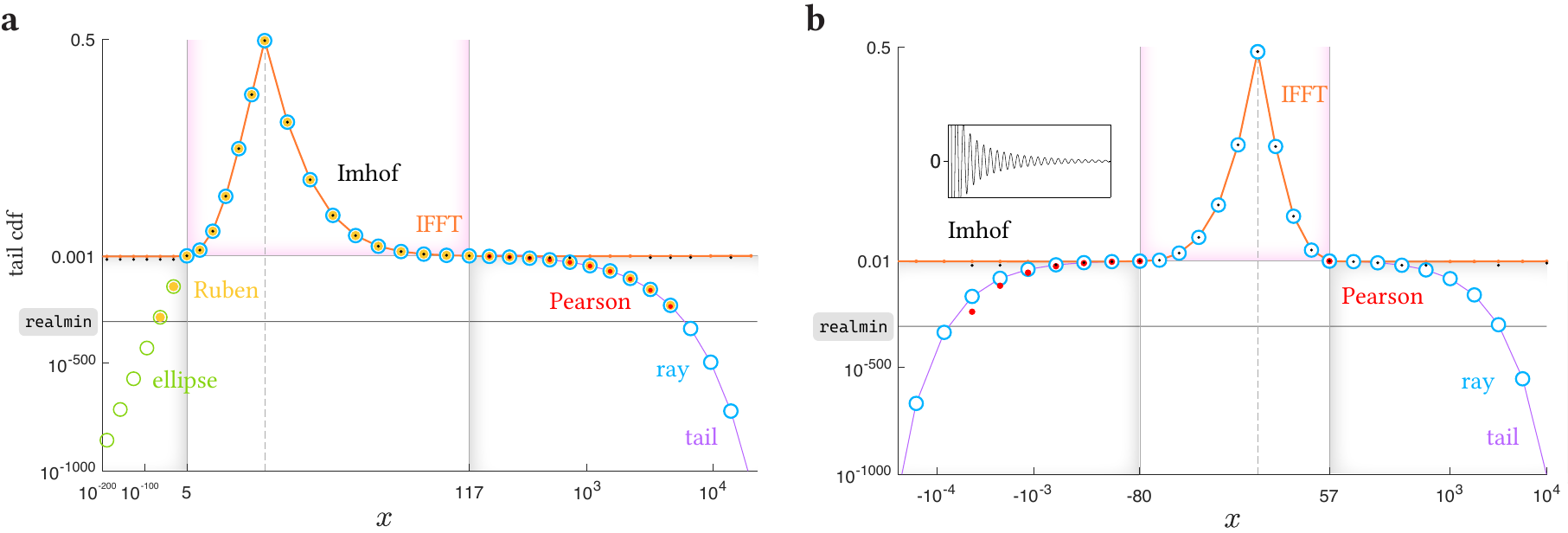}
    \end{subfigure}
    
    \vspace{1em}
    
    \begin{subfigure}[b]{0.45\textwidth}
        \hspace{2.6em}
        \resizebox{!}{6em}{
        {\tabulinesep=0.7mm
        \begin{tabu}{|l|l|l|c|}
        \hline
            \multirow{2}{*}{method} & \multicolumn{2}{c|}{smallest prob.} & \multirow{2}{*}{time/point} \\
            \cline{2-3}
            & lower & upper & \\
        \hline
        Imhof & $10^{-3}$ & $10^{-3}$ & 20 ms (dp), 0.5 s (vp)\\
        \textcolor{myorange}{IFFT} & $10^{-3}$ & $10^{-3}$ & 60 ms\\
        \textcolor{mysky}{ray} & $10^{-3}$ & $10^{-10^{308}}$ & 0.2 s (GPU)\\       \textcolor{myyellow}{Ruben} & $10^{-308}$ & $10^{-308}$ & 2 ms (lower), 0.2 s (upper)\\
        \textcolor{mygreen}{ellipse} & $10^{-10^{308}}$ & --- & 0.2 ms\\
        \textcolor{red}{Pearson} & --- & $10^{-308}$ & 0.1 ms\\
        \textcolor{mypurple}{tail} & --- & $10^{-10^{308}}$ & 0.1 ms\\
        \hline
        \end{tabu}}}
    \end{subfigure}
    \hfill
    \begin{subfigure}[b]{0.45\textwidth}
        \hspace{1.5em}
        \resizebox{!}{5em}{
        {\tabulinesep=0.7mm
        \begin{tabu}{|l|l|l|c|}
        \hline
            \multirow{2}{*}{method} & \multicolumn{2}{c|}{smallest prob.} & \multirow{2}{*}{time/point} \\
            \cline{2-3}
            & lower & upper & \\
        \hline
        Imhof & $10^{-2}$ & $10^{-2}$ & 4 ms (dp), 1.3 s (vp)\\
        \textcolor{myorange}{IFFT} & $10^{-2}$ & $10^{-2}$ & 70 ms\\
        \textcolor{mysky}{ray} & $10^{-10^{308}}$ & $10^{-10^{308}}$ & 0.2 s (GPU)\\ 
        \textcolor{red}{Pearson} & $10^{-308}$ & $10^{-3}$ & 0.5 ms\\
        \textcolor{mypurple}{tail} & $10^{-10^{308}}$ & $10^{-10^{308}}$ & 1.5 ms\\
        \hline
        \end{tabu}}}
    \end{subfigure}
    \caption{Computing the generalized chi-square cdf. \textbf{a.} Cdf of a generalized chi-square with a lower finite and upper infinite tail, computed with several methods. We plot the lower tail probability (cdf) till the median (dashed vertical line), and upper tail probability (ccdf) beyond it. The middle of the distribution, probabilities ${>}0.001$ (area highlighted pink) is in linear axes, and the tail regions (areas highlighted grey) are in double log axes. $\code{realmin}=10^{-308}$ is the double-precision limit. Table below shows the orders of the smallest lower and upper tail probabilities reached by the methods here, and their computation times per point. `dp' and `vp' mean double and variable precision. \textbf{b.} Cdf of a generalized chi-square with two infinite tails computed with several methods. Middle area of probability ${>}0.01$ is in linear axes, tail areas are in double log axes. Inset: Imhof integrand at $x=-2000$. Similar table below.}
    \label{fig:5_cdf}
\end{figure*}

In fig. \ref{fig:5_cdf}a, we compute the cdf of a generalized chi-square with $\bm{w} = \begin{bmatrix} 2 & 4 & .5 \end{bmatrix}$, $\bm{k} = \begin{bmatrix} 3 & 5 & 1 \end{bmatrix}$, $\bm{\lambda} = \begin{bmatrix} 4 & 1 & .3 \end{bmatrix}$, $s=m=0$, which has a lower finite tail and an upper infinite tail. We use Imhof's method (with double precision for the middle and variable precision in the tails, and a relative error tolerance of $10^{-10}$), IFFT (with a grid of $N=10^7$ points across a span of $10^7$), ray-tracing (with $10^6$ rays on the GPU, and the \code{log} precision setting, which automatically switches to the log approximation below \code{realmin}), Ruben's method (with 100 terms in the lower tail and $10^4$ in the upper). In only the lower tail (left region highlighted grey) we use the ellipse approximation, and in the upper tail (right region highlighted grey) we use the Pearson approximation, and infinite-tail approximation, called simply `tail' from now on (with automatic log approximation below \code{realmin}).

To clearly show both the center and the tails of the distribution in the same plot, we have used a few plotting tricks. First, we plot the lower tail probability (cdf) until the median (cdf=0.5), and the upper tail probability (complementary cdf) beyond it, so that we can show both tails on the same vertical log axes. Second, we select the main body of the distribution here to be the region above a lower or upper tail probability of 0.001 (area highlighted pink). We use linear horizontal and vertical axes to show this area clearly, but in the tails we use log horizontal and vertical axes (areas highlighted grey). Missing dots are where a method wrongly computes a small tail probability as 0, so it cannot be shown on a log axis.

Imhof's method reaches accurately down to about $10^{-3}$ in both the lower and upper tails, beyond which it returns incorrect values, sometimes negative, which we clip to 0, so those points are missing on this log plot. The smallest cdf or pdf values reachable using Imhof's method with double or variable precision varies greatly with the distribution, and is not very consistently predictable. We shall elaborate on its limitations more while discussing fig. \ref{fig:5_cdf}b. The IFFT method with its settings here also levels out at around $10^{-3}$ in both tails, so it is not very accurate without sacrificing much more in terms of speed. Ruben's method computes accurately down to about \code{realmin} in both tails. The ellipse method can compute accurately all the way down to $10^{-10^{308}}$ in the finite lower tail. The ray-trace method stops at $10^{-3}$ in the lower tail and returns 0 below that, due to the reason of the vanishing ellipsoidal integration domain we described in sec. \ref{sec:ray-tracing}. This limits the ray method in this finite tail well before \code{realmin}, so \code{log} or \code{vpa} precision does not even kick in. In the upper tail, it reaches $\code{realmin}$, beyond which it uses the log approximation to continue down to $10^{-10^{308}}$. The Pearson approximation computes down to  \code{realmin} in the upper tail, but starts dipping slightly below the other methods in this example. The tail approximation reaches \code{realmin} in the upper tail, and continues without any issues down to $10^{-10^{308}}$ using the log scale.

In the table below the figure, we list the smallest upper and lower tail probabilities that could be reached using these methods with these settings, along with the computation speeds.

In summary, we see that in the center of the distribution, all the exact methods agree well. In the finite lower tail, ellipse method agrees with Ruben and continues down into the far tail. In the infinite upper tail, Pearson is a bit erroneous, while ray and tail methods are consistent as they continue into the far tail.

We can also use our methods to compute the cdf or pdf of a non-central chi-square, which is a special case of the generalized chi-square. Matlab's \code{ncx2cdf} and \code{ncx2pdf} implement a fast and relatively straightforward calculation for this case, and they can reach down to \code{realmin} in the finite lower tail, where the ray method again stops short due to the vanishing domain. In the infinite upper tail, \code{ncx2pdf} stops at \code{realmin}, but \code{ncx2cdf} may stop much sooner. Ray-trace can reach \code{realmin} there with double precision, and $10^{-10^{308}}$ using the log scale. So for non-central chi-square cdf's, \code{ncx2cdf} and \code{ncx2pdf} are the best method for the lower tail and body of the distribution, the ellipse method is again good for the lower tail, and \code{ncx2pdf}, ray, and infinite-tail methods are best for the upper tail.

In fig. \ref{fig:5_cdf}b we compute probabilities of a generalized chi-square with $\bm{w} = \begin{bmatrix} 1 & -5 & 2 \end{bmatrix}$, $\bm{k} = \begin{bmatrix} 1 & 2 & 3 \end{bmatrix}$, $\bm{\lambda} = \begin{bmatrix} 2 & 3 & 7 \end{bmatrix}$, $s=10$ and $m=5$, which has two infinite tails. We use Imhof, IFFT, ray, Pearson and tail methods with the same settings as before. In the center of the distribution the exact methods again agree well. But in the tails, the Imhof method again computes small probabilities inaccurately, and they are sometimes negative (which we clip to 0 and these are the missing black dots). This is because in the tail, the Imhof integrand related to the characteristic function (see inset) is highly oscillatory and integrates to a tiny probability. So when numerically integrated, the larger oscillations must be computed to precisely cancel and accurately produce the tiny residual sum. When this does not work so precisely, the integral is either above or below the true value. The ray method is an entirely different approach where the contribution to the integral from each ray is non-negative, and they never cancel each other. So it avoids this problem and can again reach down to \code{realmin}, and then to $10^{-10^{308}}$ using the log approximation, with fast double precision, in both tails here because they are both infinite. The IFFT method is fast, but less capable of accurately reaching small tail probabilities. The Pearson approximation here reaches \code{realmin} in the lower tail, but has worse errors, whereas in the upper tail it stops much shorter. The tail approximation can reach \code{realmin} in both tails, and continues to $10^{-10^{308}}$ using the log scale without any errors.

The table below the figure lists the smallest tail probabilities that could be reached by these methods with these settings, along with their speeds. In summary, the exact methods again agree well in the center of the distribution, whereas in the infinite tails, ray and tail methods are the best.

To compute the inverse cdf, our \code{gx2inv} function uses numerical root-finding on the forward cdf function, which can automatically pick the best cdf method given the distribution parameters. Since the ellipse method can compute tiny probabilities in finite tails, and ray and tail methods can do so in infinite tails, \code{gx2inv} can leverage them to find quantiles corresponding to miniscule tail probabilities, which is not possible by inverting other cdf methods.

\subsection{Computing the pdf}
\begin{figure*}[ht]
    \centering
    \begin{subfigure}[b]{\textwidth}
        \centering
        \includegraphics[width=\linewidth]{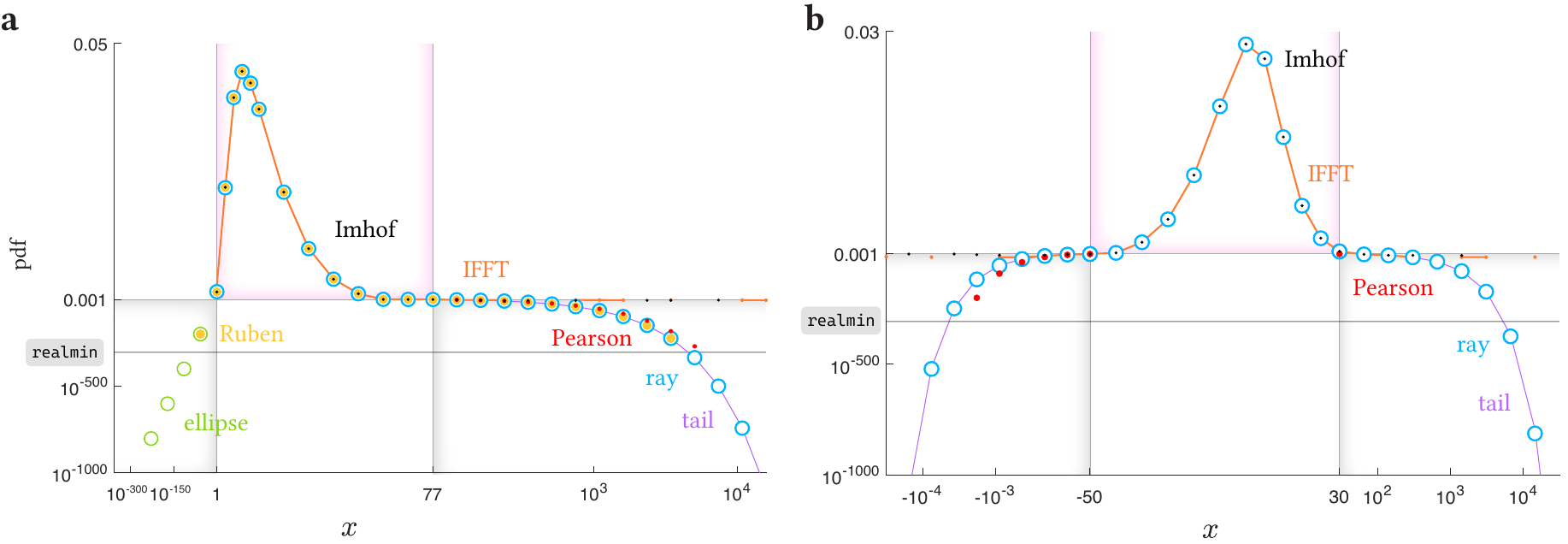}
    \end{subfigure}
    
    \vspace{1em}
    
    \begin{subfigure}[b]{0.45\textwidth}
        \hspace{2.6em}
        \resizebox{!}{6em}{
        {\tabulinesep=0.7mm
        \begin{tabu}{|l|l|l|c|}
        \hline
            \multirow{2}{*}{method} & \multicolumn{2}{c|}{smallest density} & \multirow{2}{*}{time/point} \\
            \cline{2-3}
            & lower & upper & \\
        \hline
        Imhof & $10^{-3}$ & $10^{-4}$ & 80 ms (dp), 4 s (vp)\\
        \textcolor{myorange}{IFFT} & $10^{-2}$ & $10^{-4}$ & 70 ms\\
        \textcolor{mysky}{ray} & $10^{-3}$ & $10^{-10^{308}}$ & 1.2 s (GPU)\\
        \textcolor{myyellow}{Ruben} & $10^{-308}$ & $10^{-308}$ & 10 ms (lower), 0.8 s (upper)\\
        \textcolor{mygreen}{ellipse} & $10^{-10^{308}}$ & --- & 0.5 ms\\
        \textcolor{red}{Pearson} & --- & $10^{-308}$ & 0.1 ms\\
        \textcolor{mypurple}{tail} & --- & $10^{-10^{308}}$ & 0.3 ms\\
        \hline
        \end{tabu}}}
    \end{subfigure}
    \hfill
    \begin{subfigure}[b]{0.45\textwidth}
        \hspace{1.5em}
        \resizebox{!}{5em}{
        {\tabulinesep=0.7mm
        \begin{tabu}{|l|l|l|c|}
        \hline
            \multirow{2}{*}{method} & \multicolumn{2}{c|}{smallest density} & \multirow{2}{*}{time/point} \\
            \cline{2-3}
            & lower & upper & \\
        \hline
        Imhof & $10^{-3}$ & $10^{-3}$ & 30 ms (dp), 0.9 s (vp)\\
        \textcolor{myorange}{IFFT} & $10^{-3}$ & $10^{-3}$ & 80 ms\\
        \textcolor{mysky}{ray} & $10^{-10^{308}}$ & $10^{-10^{308}}$ & 0.15 s (GPU)\\ 
        \textcolor{red}{Pearson} & $10^{-308}$ & $10^{-3}$ & 0.4 ms\\
        \textcolor{mypurple}{tail} & $10^{-10^{308}}$ & $10^{-10^{308}}$ & 0.2 ms\\
        \hline
        \end{tabu}}}
    \end{subfigure}
    \caption{Computing the generalized chi-square pdf. \textbf{a.} Pdf of a generalized chi-square with a lower finite and upper infinite tail, computed with several methods. The middle of the distribution, probabilities ${>}0.001$ (area highlighted pink) is in linear axes, tail regions (highlighted grey) are in double log axes. The table below shows the rough orders of the smallest densities computed accurately in the lower and upper tails here, and their computation times per point. \textbf{b.} Pdf of a generalized chi-square with two infinite tails computed with several methods. Middle area of probability ${>}0.001$ is in linear axes, tail areas are in log axes. Similar table below.}
    \label{fig:6_pdf}
\end{figure*}

In fig. \ref{fig:6_pdf}a, we compute the pdf of a generalized chi-square with $\bm{w} = \begin{bmatrix} 1 & 3 & .5 & .2 \end{bmatrix}$, $\bm{k} = \begin{bmatrix} 3 & 1 & 2 & 1 \end{bmatrix}$, $\bm{\lambda} = \begin{bmatrix} 0 & 3 & 5 & 0 \end{bmatrix}$, $s=m=0$, which has a lower finite tail and an upper infinite tail. We use Imhof, IFFT and Ruben's methods with the same settings as for the cdf, ray-tracing (on the GPU, and with $10^7$ rays here for greater accuracy), the ellipse approximation (only in the lower tail), and Pearson and tail approximations (only in the upper tail).

We see again that in the center of the distribution, all the exact methods agree well. Imhof and IFFT methods reach accurately down to only about $10^{-3}$ in both tails. Ruben's method again reaches accurately down to about \code{realmin} in both tails. The ellipse method can again compute accurately to $10^{-10^{308}}$ in the lower tail. Ray-trace stops at $10^{-3}$ in the lower tail (log or variable precision does not help here again), but in the upper tail, it can reach $\code{realmin}$, and then $10^{-10^{308}}$ using the log approximation. Pearson's approximation can reach \code{realmin} in the upper tail, but is erroneous, whereas the tail approximation continues to \code{realmin}, and then to $10^{-10^{308}}$ using the log approximation, without any errors.

In the table below the figure, we list the smallest probability densities computed correctly in the lower and upper tails using these methods with their settings, along with their computation speeds.

In fig. \ref{fig:6_pdf}b we compute probabilities of a generalized chi-square with $\bm{w} = \begin{bmatrix} 4 & -1 & 2 & -3\end{bmatrix}$, $\bm{k} = \begin{bmatrix} 1 & 1 & 2 & 3 \end{bmatrix}$, $\bm{\lambda} = \begin{bmatrix} 0 & 4 & 0 & 2 \end{bmatrix}$, $s=3$ and $m=10$, which has two infinite tails. We use Imhof's method (with double precision and relative tolerance of $10^{-10}$, and variable precision and relative tolerance of only $10^{-1}$, because no finer tolerance could be reached), IFFT (with the same settings as before), and ray-trace (with $10^6$ rays on the GPU). In the center of the distribution the methods again agree well. But in the tails, Imhof and IFFT stop early, whereas the ray method can again reach down to $10^{-10^{308}}$ pretty fast. The Pearson approximation reaches \code{realmin} in the lower tail, but has significant errors, whereas in the upper tail it again stops much shorter. The tail approximation can reach $10^{-10^{308}}$ in both tails without any errors. 

In summary, just as with the cdf, the exact methods again agree well in the center of the distribution. Ruben and ellipse work well in the finite tail, and ray and tail methods are the best in the infinite tail.

Based on our comparisons, we list in table \ref{tab:1_best_methods} the cdf and pdf methods discussed in this paper that work the best in different parts of a generalized chi-square distribution in different cases.

\begin{table}
    \centering
\resizebox{\columnwidth}{!}{    
    {\tabulinesep=1.4mm
        \begin{tabu}{|c|l|l|}
        \hline
        \textbf{$\tilde{\chi} \ $ type} &  \textbf{part} & \textbf{best cdf/pdf method(s)}\\
        \hline
        \multirow{3}{*}{\makecell{ellipse: \\ $w_i$ same sign, \\ $s=0$}} & body & Ruben, Imhof, IFFT, ray\\
        & finite tail & Ruben, ray (if $\lambda_i=0$), ellipse \\
        & infinite tail & Ruben, ray, tail\\
        \hline
        \multirow{2}{*}{\makecell{not ellipse: \\$w_i$ mixed signs, \\ and/or $s \neq 0$}} & body & Imhof, IFFT, ray\\
        & infinite tails & ray, tail\\
        \hline
        \multirow{3}{*}{\makecell{sphere: \\non-central $\chi^2$ \\(only one term)}} & body & \code{ncx2cdf}/\code{ncx2pdf}\\
        & finite tail & \code{ncx2cdf}/\code{ncx2pdf}, ellipse\\
        & infinite tail & \code{ncx2pdf}, ray, tail\\
        \hline
    \end{tabu}}}
    \caption{The best cdf and pdf methods to use for the different parts of the generalized chi-square distribution in different cases. `Tail' refers to the infinite-tail approximation.}
    \label{tab:1_best_methods}
\end{table}

\section{More accuracy tests}

\subsection{Against published tables}

\begin{table*}[!h]
\begin{center}
\resizebox{\textwidth}{!}{
\begin{tabular}{|r|l|r|l|l|c|l|c|l|c|}
\hline
    \multirow{2}{*}{\#} & \multirow{2}{*} {$\tilde{\chi} \ $ parameters} & \multirow{2}{*}{$x$} & \multirow{2}{*}{exact value} & \multicolumn{2}{c|}{Imhof}  & \multicolumn{2}{c|}{ray} & \multicolumn{2}{c|}{IFFT}\\
    \cline{5-10}
    & & & & value & time & value & time & value & time\\
\hline

\multirow{3}{*}{1} & \multirow{3}{*}{
\scriptsize{$\!\begin{aligned}[t]
    \bm{w} &= \begin{bmatrix}.6 & .3 & .1 \end{bmatrix} \\
    \bm{k} &= \begin{bmatrix}1 & 1 & 1 \end{bmatrix} \\
    \bm{\lambda}&= \begin{bmatrix}0 & 0 & 0 \end{bmatrix}
    \end{aligned}$}
} & .1 & \cellcolor{mylightgreen} .9458 & \cellcolor{mylightgreen} .9458(1) & \multirow{3}{*}{0.04 s} & \cellcolor{mylightgreen} .9458(1) & \multirow{3}{*}{\makecell{3.5 s \\ \scriptsize{GPU, MC w $10^7$ rays}}} & \cellcolor{mylightgreen}{.9458} & \multirow{3}{*}{\makecell{0.1 s \\ \scriptsize{$ \Delta x=10^4, N=10^6$}}} \\
\cline{3-5} \cline{7-7} \cline{9-9}
 & & .7 & \cellcolor{mylightgreen} .5064 & \cellcolor{mylightgreen} .5064(1) & & \cellcolor{mylightgreen}{.5064(1)} & & \cellcolor{mylightgreen}{.5064} & \\
\cline{3-5} \cline{7-7} \cline{9-9}
 & & 2 & \cellcolor{mylightgreen} .1240 & \cellcolor{mylightgreen} .1240(1) & & \cellcolor{mylightgreen}{.1240(1)} & & \cellcolor{mylightgreen}{.1240} & \\
\hline

\multirow{3}{*}{2} & \multirow{3}{*}{
\scriptsize{$\!\begin{aligned}[t]
    \bm{w} &= \begin{bmatrix}.6 & .3 & .1 \end{bmatrix} \\
    \bm{k} &= \begin{bmatrix}2 & 2 & 2 \end{bmatrix} \\
    \bm{\lambda}&= \begin{bmatrix}0 & 0 & 0 \end{bmatrix}
    \end{aligned}$}
} & .2 & .9936 & \cellcolor{mylightgreen} .993547(1)  & \multirow{3}{*}{0.03 s} & \cellcolor{mylightgreen} .993545(5) & \multirow{3}{*}{\makecell{3.5 s \\ \scriptsize{GPU, MC w $10^7$ rays}}} & \cellcolor{mylightgreen} .993546 & \multirow{3}{*}{\makecell{9.5 s \\ \scriptsize{$ \Delta x=10^6, N=10^8$}}} \\
\cline{3-5} \cline{7-7} \cline{9-9}
& & 2 & \cellcolor{mylightgreen} .3998 & \cellcolor{mylightgreen}.399795(1) & & \cellcolor{mylightgreen}.3997(1) & & \cellcolor{mylightgreen}.3998 & \\
\cline{3-5} \cline{7-7} \cline{9-9}
& & 6 & \cellcolor{mylightgreen} .0161 & \cellcolor{mylightgreen}.016103(1) & & \cellcolor{mylightgreen}.01610(1) & & \cellcolor{mylightgreen}.0161 & \\
\hline

\multirow{3}{*}{3} & \multirow{3}{*}{
\scriptsize{$\!\begin{aligned}[t]
    \bm{w} &= \begin{bmatrix}.6 & .3 & .1 \end{bmatrix} \\
    \bm{k} &= \begin{bmatrix}6 & 4 & 2 \end{bmatrix} \\
    \bm{\lambda}&= \begin{bmatrix}0 & 0 & 0 \end{bmatrix}
    \end{aligned}$}
} & 1 & \cellcolor{mylightgreen} .9973 & \cellcolor{mylightgreen}.9973(1) & \multirow{3}{*}{0.01 s} & \cellcolor{mylightgreen}.997319(3) & \multirow{3}{*}{\makecell{4.7 s \\ \scriptsize{GPU, MC w $10^7$ rays}}} & \cellcolor{mylightgreen}.9973 & \multirow{3}{*}{\makecell{0.1 s \\ \scriptsize{$ \Delta x=10^5, N=10^6$}}} \\
\cline{3-5} \cline{7-7} \cline{9-9}
& & 5 & \cellcolor{mylightgreen} .4353 & \cellcolor{mylightgreen}.4353(1) & & \cellcolor{mylightgreen}.4353(1) & & \cellcolor{mylightgreen}{.4353} & \\
\cline{3-5} \cline{7-7} \cline{9-9}
& & 12 & \cellcolor{mylightgreen} .0088 & \cellcolor{mylightgreen}.0088(1) & & \cellcolor{mylightgreen}.008770(7) & & \cellcolor{mylightgreen}.0088 & \\
\hline

\multirow{3}{*}{4} & \multirow{3}{*}{
\scriptsize{$\!\begin{aligned}[t]
    \bm{w} &= \begin{bmatrix}.6 & .3 & .1 \end{bmatrix} \\
    \bm{k} &= \begin{bmatrix}2 & 4 & 6 \end{bmatrix} \\
    \bm{\lambda}&= \begin{bmatrix}0 & 0 & 0 \end{bmatrix}
    \end{aligned}$}
} & 1 & \cellcolor{mylightgreen} .9666 & \cellcolor{mylightgreen}.9666(1) & \multirow{3}{*}{8 ms} & \cellcolor{mylightgreen}.96665(3) & \multirow{3}{*}{\makecell{4.8 s \\ \scriptsize{GPU, MC w $10^7$ rays}}} & \cellcolor{mylightgreen}.9666 & \multirow{3}{*}{\makecell{0.08 s \\ \scriptsize{$ \Delta x=10^5, N=10^6$}}} \\
\cline{3-5} \cline{7-7} \cline{9-9}
& & 3 & \cellcolor{mylightgreen} .4196 & \cellcolor{mylightgreen}.4196(1) & & \cellcolor{mylightgreen}.4196(1) & & \cellcolor{mylightgreen}.4196 & \\
\cline{3-5} \cline{7-7} \cline{9-9}
& & 8 & \cellcolor{mylightgreen} .0087 & \cellcolor{mylightgreen}.00871(1) & & \cellcolor{mylightgreen}.00871(1) & & \cellcolor{mylightgreen}.0087 & \\
\hline

\multirow{3}{*}{5} & \multirow{3}{*}{
\scriptsize{$\!\begin{aligned}[t]
    \bm{w} &= \begin{bmatrix}.7 & .3 \end{bmatrix} \\
    \bm{k} &= \begin{bmatrix}6 & 2 \end{bmatrix} \\
    \bm{\lambda}&= \begin{bmatrix}6 & 2 \end{bmatrix}
    \end{aligned}$}
} & 2 & \cellcolor{mylightgreen} .9939 & \cellcolor{mylightgreen}.9939(1) & \multirow{3}{*}{7 ms} & \cellcolor{mylightgreen}.99391(9) & \multirow{3}{*}{\makecell{1 s \\ \scriptsize{GPU, MC w $10^6$ rays}}} & \cellcolor{mylightgreen}.9939 & \multirow{3}{*}{\makecell{1 s \\ \scriptsize{$ \Delta x=10^6, N=10^7$}}} \\
\cline{3-5} \cline{7-7} \cline{9-9}
& & 10 & \cellcolor{mylightgreen} .4087 & \cellcolor{mylightgreen}.4087(1) & & \cellcolor{mylightgreen}.4087(2) & & \cellcolor{mylightgreen}.4087 & \\
\cline{3-5} \cline{7-7} \cline{9-9}
& & 20 & \cellcolor{mylightgreen} .0221 & \cellcolor{mylightgreen}.0221(1) & & \cellcolor{mylightgreen}.02210(6) & & \cellcolor{mylightgreen}.0221 & \\
\hline

\multirow{3}{*}{6} & \multirow{3}{*}{
\scriptsize{$\!\begin{aligned}[t]
    \bm{w} &= \begin{bmatrix}.7 & .3 \end{bmatrix} \\
    \bm{k} &= \begin{bmatrix}1 & 1 \end{bmatrix} \\
    \bm{\lambda}&= \begin{bmatrix}6 & 2 \end{bmatrix}
    \end{aligned}$}
} & 1 & \cellcolor{mylightgreen} .9549 & \cellcolor{mylightgreen}.9549(1) & \multirow{3}{*}{0.01 s} & \cellcolor{mylightgreen}.9548(1) & \multirow{3}{*}{\makecell{0.1 s \\ \scriptsize{grid integral}}} & \cellcolor{mylightgreen}.9549 & \multirow{3}{*}{\makecell{1 s \\ \scriptsize{$ \Delta x=10^6, N=10^7$}}} \\
\cline{3-5} \cline{7-7} \cline{9-9}
& & 6 & \cellcolor{mylightgreen} .4076 & \cellcolor{mylightgreen}.4076(1) & & \cellcolor{mylightgreen}.4076(1) & & \cellcolor{mylightgreen}.4076 & \\
\cline{3-5} \cline{7-7} \cline{9-9}
& & 15 & \cellcolor{mylightgreen} .0223 & \cellcolor{mylightgreen}.0223(1) & & \cellcolor{mylightgreen}{.0223(1)} & & \cellcolor{mylightgreen}.0223 & \\
\hline

\multirow{3}{*}{7} & \multirow{3}{*}{
\scriptsize{$\!\begin{aligned}[t]
    \bm{w} &= \begin{bmatrix}.2 & .1 & \frac{.1}{3} & .4 & \frac{.2}{3}\end{bmatrix} \\
    \bm{k} &= \begin{bmatrix}10 & 4 & 2 & 2 & 6 \end{bmatrix} \\
    \bm{\lambda}&= \begin{bmatrix}0 & 0 & 0 & 0 & 0 \end{bmatrix}
    \end{aligned}$}
} & 1.5 & \cellcolor{mylightgreen} .9891 & \cellcolor{mylightgreen}.9891(1) & \multirow{3}{*}{0.01 s} & \cellcolor{mylightgreen}.98906(1) & \multirow{3}{*}{\makecell{8.4 s \\ \scriptsize{GPU, MC w $10^7$ rays}}} & \cellcolor{mylightgreen}.9891 & \multirow{3}{*}{\makecell{2.5 s \\ \scriptsize{$ \Delta x=2 {\times} 10^5, N=2 {\times} 10^7$}}} \\
\cline{3-5} \cline{7-7} \cline{9-9}
& & 4 & \cellcolor{mylightgreen} .3453 & \cellcolor{mylightgreen}.3453(1) & & \cellcolor{mylightgreen}.3453(1) & & \cellcolor{mylightgreen}.3453 & \\
\cline{3-5} \cline{7-7} \cline{9-9}
& & 7 & \cellcolor{mylightgreen} .0154 & \cellcolor{mylightgreen}.0154(1) & & \cellcolor{mylightgreen}.01541(2) & & \cellcolor{mylightgreen}.0154 & \\
\hline

\multirow{3}{*}{8} & \multirow{3}{*}{
\scriptsize{$\!\begin{aligned}[t]
    \bm{w} &= \begin{bmatrix}.2 & .1 & \frac{.1}{3} & -.4 & -.2 & -\frac{.2}{3}\end{bmatrix} \\
    \bm{k} &= \begin{bmatrix}6 & 4 & 2 & 2 & 4 & 6 \end{bmatrix} \\
    \bm{\lambda}&= \begin{bmatrix}0 & 0 & 0 & 0 & 0 & 0 \end{bmatrix}
    \end{aligned}$}
} & -2 & \cellcolor{mylightgreen} .9102 & \cellcolor{mylightgreen}.910225(1) & \multirow{3}{*}{.01 s} & \cellcolor{mylightgreen}.91024(4) & \multirow{3}{*}{\makecell{2.8 m \\ \scriptsize{GPU, MC w $2{\times}10^8$ rays}}} & \cellcolor{mylightgreen}.9102 & \multirow{3}{*}{\makecell{14 s \\ \scriptsize{$ \Delta x=10^6, N=10^8$}}} \\
\cline{3-5} \cline{7-7} \cline{9-9}
& & 0 & \cellcolor{mylightgreen} .4061 & \cellcolor{mylightgreen}.406106(1) & & \cellcolor{mylightgreen}.40609(9) & & \cellcolor{mylightgreen}.4061 & \\
\cline{3-5} \cline{7-7} \cline{9-9}
& & 2.5 & .0097 & \cellcolor{mylightgreen} .009760(1) & & \cellcolor{mylightgreen} .009758(7) & & \cellcolor{mylightgreen} .00976 & \\
\hline

\multirow{3}{*}{9} & \multirow{3}{*}{
\scriptsize{$\!\begin{aligned}[t]
    \bm{w} &= \begin{bmatrix}\frac{.7}{2} & \frac{.3}{2} \end{bmatrix} \\
    \bm{k} &= \begin{bmatrix}7 & 3 \end{bmatrix} \\
    \bm{\lambda}&= \begin{bmatrix}12 & 4 \end{bmatrix}
    \end{aligned}$}
} & 3.5 & \cellcolor{mylightgreen} .9563 & \cellcolor{mylightgreen}.9563(1) & \multirow{3}{*}{7 ms} & \cellcolor{mylightgreen}.9564(2) & \multirow{3}{*}{\makecell{0.7 s \\ \scriptsize{GPU, MC w $10^6$ rays}}} & \cellcolor{mylightgreen}.9563 & \multirow{3}{*}{\makecell{9.5 s \\ \scriptsize{$ \Delta x=10^6, N=10^8$}}} \\
\cline{3-5} \cline{7-7} \cline{9-9}
& & 8 & \cellcolor{mylightgreen} .4152 & \cellcolor{mylightgreen}.4152(1) & & \cellcolor{mylightgreen}.4152(1) & & \cellcolor{mylightgreen}.4152 & \\
\cline{3-5} \cline{7-7} \cline{9-9}
& & 13 & \cellcolor{mylightgreen} .0462 & \cellcolor{mylightgreen}.0462(1) & & \cellcolor{mylightgreen}.0463(1) & & \cellcolor{mylightgreen}.0462 & \\
\hline

\multirow{3}{*}{10} & \multirow{3}{*}{
\scriptsize{$\!\begin{aligned}[t]
    \bm{w} &= \begin{bmatrix}\frac{.7}{2} & \frac{.3}{2} & -\frac{.7}{2} & -\frac{.3}{2}\end{bmatrix} \\
    \bm{k} &= \begin{bmatrix}6 & 2 & 1 & 1 \end{bmatrix} \\
    \bm{\lambda}&= \begin{bmatrix}6 & 2 & 6 & 2 \end{bmatrix}
    \end{aligned}$}
} & -2 & \cellcolor{mylightgreen} .9218 & \cellcolor{mylightgreen}.9218(1) & \multirow{3}{*}{9 ms} & \cellcolor{mylightgreen}.9215(3) & \multirow{3}{*}{\makecell{0.8 s \\ \scriptsize{GPU, MC w $10^6$ rays}}} & \cellcolor{mylightgreen}.9218 & \multirow{3}{*}{\makecell{1.2 s \\ \scriptsize{$ \Delta x=10^6, N=10^7$}}} \\
\cline{3-5} \cline{7-7} \cline{9-9}
& & 2 & \cellcolor{mylightgreen} .4779 & \cellcolor{mylightgreen}.4779(1) & & \cellcolor{mylightgreen}.4780(2) & & \cellcolor{mylightgreen}.4779 & \\
\cline{3-5} \cline{7-7} \cline{9-9}
& & 7 & \cellcolor{mylightgreen} .0396 & \cellcolor{mylightgreen}.0396(1) & & \cellcolor{mylightgreen}.0397(1) & & \cellcolor{mylightgreen}.0396 & \\
\hline

\multirow{3}{*}{11} & \multirow{3}{*}{
\scriptsize{$\!\begin{aligned}[t]
    \bm{w} &= \begin{bmatrix}\frac{.6}{4} & \frac{.3}{4} & \frac{.1}{4} & \frac{.7}{4}\end{bmatrix} \\
    \bm{k} &= \begin{bmatrix}8 & 11 & 8 & 7 \end{bmatrix} \\
    \bm{\lambda}&= \begin{bmatrix}0 & 4 & 0 & 12 \end{bmatrix}
    \end{aligned}$}
} & 3 & \cellcolor{mylightgreen} .9842 & \cellcolor{mylightgreen}.9842(1) & \multirow{3}{*}{6 ms} & \cellcolor{mylightgreen}.9842(1) & \multirow{3}{*}{\makecell{1.1 s \\ \scriptsize{GPU, MC w $10^6$ rays}}} & \cellcolor{mylightgreen}.9842 & \multirow{3}{*}{\makecell{1.3 s \\ \scriptsize{$ \Delta x=10^6, N=10^7$}}} \\
\cline{3-5} \cline{7-7} \cline{9-9}
& & 6 & \cellcolor{mylightgreen} .4264 & \cellcolor{mylightgreen}.4264(1) & & \cellcolor{mylightgreen}.4263(3) & & \cellcolor{mylightgreen}.4264 & \\
\cline{3-5} \cline{7-7} \cline{9-9}
& & 10 & \cellcolor{mylightgreen} .0117 & \cellcolor{mylightgreen}.0117(1) & & \cellcolor{mylightgreen}.01165(6) & & \cellcolor{mylightgreen}.0117 & \\
\hline

\multirow{3}{*}{12} & \multirow{3}{*}{
\scriptsize{$\!\begin{aligned}[t]
    \bm{w} &= \begin{bmatrix}.1 & \frac{.1}{2} & \frac{.1}{6} & -\frac{.7}{6} -\frac{.1}{2} & \frac{.7}{3} & -.2 & -.1 & -\frac{.1}{3}\end{bmatrix} \\
    \bm{k} &= \begin{bmatrix}7 & 4 & 2 & 6 & 2 & 1 & 2 & 4 & 6 \end{bmatrix} \\
    \bm{\lambda}&= \begin{bmatrix}2 & 0 & 0 & 6 & 2 & 6 & 0 & 0 & 0 \end{bmatrix}
    \end{aligned}$}
} & -3 & \cellcolor{mylightgreen} .9861 & \cellcolor{mylightgreen}.9861(1) & \multirow{3}{*}{8 ms} & \cellcolor{mylightgreen}.9862(1) & \multirow{3}{*}{\makecell{1.1 s \\ \scriptsize{GPU, MC w $10^6$ rays}}} & \cellcolor{mylightgreen}.9861 & \multirow{3}{*}{\makecell{0.1 s \\ \scriptsize{$ \Delta x=10^5, N=10^6$}}} \\
\cline{3-5} \cline{7-7} \cline{9-9}
& & 0 & \cellcolor{mylightgreen} .5170 & \cellcolor{mylightgreen}.5170(1) & & \cellcolor{mylightgreen}.5170(5) & & \cellcolor{mylightgreen}.5170 & \\
\cline{3-5} \cline{7-7} \cline{9-9}
& & 4 & \cellcolor{mylightgreen} .0152 & \cellcolor{mylightgreen}.0152(1) & & \cellcolor{mylightgreen}.0152(1) & & \cellcolor{mylightgreen}.0152 & \\
\hline

\multirow{3}{*}{13} & \multirow{3}{*}{
\scriptsize{$\!\begin{aligned}[t]
    \bm{w} &= \begin{bmatrix}.5 & .4 & .1 \end{bmatrix} \\
    \bm{k} &= \begin{bmatrix}1 & 2 & 1 \end{bmatrix} \\
    \bm{\lambda}&= \begin{bmatrix}1 & .6 & .8 \end{bmatrix}
    \end{aligned}$}
} & 2 & \cellcolor{mylightgreen} .457461(1) & \cellcolor{mylightgreen}.457461(1) & \multirow{3}{*}{0.02 s} & \cellcolor{mylightgreen}.457461(1) & \multirow{3}{*}{\makecell{2.5 s \\ \scriptsize{grid integral}}} & \cellcolor{mylightgreen}.457460 & \multirow{3}{*}{\makecell{33 s \\ \scriptsize{$ \Delta x=4{\times}10^6, N=2{\times}10^8$}}} \\
\cline{3-5} \cline{7-7} \cline{9-9}
& & 6 & \cellcolor{mylightgreen} .031109(1) & \cellcolor{mylightgreen}.031109(1) & & \cellcolor{mylightgreen}.031109(1) & & \cellcolor{mylightgreen}.031109 & \\
\cline{3-5} \cline{7-7} \cline{9-9}
& & 8 & \cellcolor{mylightgreen} .006885(1) & \cellcolor{mylightgreen}.006885(1) & & \cellcolor{mylightgreen}.006885(1) & & \cellcolor{mylightgreen}.006886 & \\
\hline

\multirow{3}{*}{14} & \multirow{3}{*}{
\scriptsize{$\!\begin{aligned}[t]
    \bm{w} &= \begin{bmatrix}.7 & .3 \end{bmatrix} \\
    \bm{k} &= \begin{bmatrix}1 & 1 \end{bmatrix} \\
    \bm{\lambda}&= \begin{bmatrix}6 & 2 \end{bmatrix}
    \end{aligned}$}
} & 1 & \cellcolor{mylightgreen} .954873(1) & \cellcolor{mylightgreen}.954873(1) & \multirow{3}{*}{0.09 s} & \cellcolor{mylightgreen}.954873(1) & \multirow{3}{*}{\makecell{0.3 s \\ \scriptsize{grid integral}}} & \cellcolor{mylightgreen}.954872 & \multirow{3}{*}{\makecell{31 s \\ \scriptsize{$ \Delta x=4{\times}10^6, N=2{\times}10^8$}}} \\
\cline{3-5} \cline{7-7} \cline{9-9}
& & 6 & \cellcolor{mylightgreen} .407565(1) & \cellcolor{mylightgreen}.407565(1) & & \cellcolor{mylightgreen}.407565(1) & & \cellcolor{mylightgreen}.407565 & \\
\cline{3-5} \cline{7-7} \cline{9-9}
& & 15 & \cellcolor{mylightgreen} .022343(1) & \cellcolor{mylightgreen}.022343(1) & & \cellcolor{mylightgreen}.022343(1) & & \cellcolor{mylightgreen}.022344 & \\
\hline

\multirow{3}{*}{15} & \multirow{3}{*}{
\scriptsize{$\!\begin{aligned}[t]
    \bm{w} &= \begin{bmatrix}.995 & .005 \end{bmatrix} \\
    \bm{k} &= \begin{bmatrix}1 & 2 \end{bmatrix} \\
    \bm{\lambda}&= \begin{bmatrix}1 & 1 \end{bmatrix}
    \end{aligned}$}
} & 2 & \cellcolor{mylightgreen} .347939(1) & \cellcolor{mylightgreen}.347939(1) & \multirow{3}{*}{0.2 s} & \cellcolor{mylightgreen}.347939(1) & \multirow{3}{*}{\makecell{0.4 s \\ \scriptsize{grid integral}}}& \cellcolor{mylightgreen}.347938 & \multirow{3}{*}{\makecell{1.3 m \\ \scriptsize{$ \Delta x=4{\times}10^6, N=10^9$}}} \\
\cline{3-5} \cline{7-7} \cline{9-9}
& & 8 & \cellcolor{mylightgreen} .033475(1) & \cellcolor{mylightgreen}.033475(1) & & \cellcolor{mylightgreen}.033475(1) & & \cellcolor{mylightgreen}.033476 & \\
\cline{3-5} \cline{7-7} \cline{9-9}
& & 12 & \cellcolor{mylightgreen} .006748(1) & \cellcolor{mylightgreen}.006748(1) & & \cellcolor{mylightgreen}.006748(1) & & \cellcolor{mylightgreen}.006749 & \\
\hline

\multirow{3}{*}{16} & \multirow{3}{*}{
\scriptsize{$\!\begin{aligned}[t]
    \bm{w} &= \begin{bmatrix}.35 & .15 & .35 & .15 \end{bmatrix} \\
    \bm{k} &= \begin{bmatrix}1 & 1 & 6 & 2 \end{bmatrix} \\
    \bm{\lambda}&= \begin{bmatrix}6 & 2 & 6 & 2 \end{bmatrix}
    \end{aligned}$}
} & 3.5 & \cellcolor{mylightgreen} .956318(1) & \cellcolor{mylightgreen}.956318(1) & \multirow{3}{*}{7 ms} & \cellcolor{mylightgreen}.956322(8) & \multirow{3}{*}{\makecell{7 m \\ \scriptsize{GPU, MC w $10^9$ rays}}} & \cellcolor{mylightgreen}.956318 & \multirow{3}{*}{\makecell{2.8 m \\ \scriptsize{$ \Delta x=10^7, N=10^9$}}} \\
\cline{3-5} \cline{7-7} \cline{9-9}
 & & 8 & \cellcolor{mylightgreen} .415239(1) & \cellcolor{mylightgreen}.415239(1) & & \cellcolor{mylightgreen}.415234(5) & & \cellcolor{mylightgreen}.415239 & \\
\cline{3-5} \cline{7-7} \cline{9-9}
 & & 13 & \cellcolor{mylightgreen} .046231(1) & \cellcolor{mylightgreen}.046231(1) & & \cellcolor{mylightgreen}.046230(4) & & \cellcolor{mylightgreen}.046231 & \\
\hline

\end{tabular}}
\end{center}
\caption{Comparing the ccdf, $p(\tilde{\chi} > x)$, computed by our implementations of Imhof, ray-trace and IFFT methods, against exact values published by Imhof \cite{imhof1961computing} and Liu et al \cite{liu2009new}. In each row, all green values match each other within uncertainty. `MC' means Monte-Carlo integration. Reported times are to compute all three cdf values of a distribution. }
\label{tab:2_comp}
\end{table*}

\begin{table}[!h]
\begin{center}
\resizebox{\columnwidth}{!}{
\begin{tabular}{|c|r|c|c|c|c|}
\hline
    \multirow{2}{*}{\#} & \multirow{2}{*}{$x$}
    & \multicolumn{2}{c|}{$\log_{10} p$}
    & \multicolumn{2}{c|}{$\log_{10} f$} \\
\cline{3-6}
    & &
    ray & tail
    & ray & tail \\
\hline

1 & 1e3
  & \cellcolor{mylightgreen} -363.430(3)
  & \cellcolor{mylightgreen} -363.431
  & \cellcolor{mylightgreen} -363.514(4)
  & \cellcolor{mylightgreen} -363.510 \\
\hline

2 & 2e3
  & \cellcolor{mylightgreen} -723.44(6)
  & \cellcolor{mylightgreen} -723.44
  & \cellcolor{mylightgreen} -723.49(4)
  & \cellcolor{mylightgreen} -723.52\\
\hline

3 & 3e3
  & \cellcolor{mylightgreen} -1079.0(4)
  & \cellcolor{mylightgreen} -1078.6
  & \cellcolor{mylightgreen} -1078.9(3)
  & \cellcolor{mylightgreen} -1078.6 \\
\hline

4 & 1e4
  & -3.667(4)e3
  & -3.62e3
  & -3.664(5)e3
  & -3.62e3\\
\hline

5 & 1e5
  & -3.0619(0)e4
  & -3.0617e4
  & -3.0619(1)e4
  & -3.0617e4\\
\hline

6 & 4e3
  & \cellcolor{mylightgreen} -1.1636(0)e3
  & \cellcolor{mylightgreen} -1.1636e3
  & \cellcolor{mylightgreen} -1.1637(0)e3
  & \cellcolor{mylightgreen} -1.1637e3\\
\hline

7 & 1e3
  & \cellcolor{mylightred} -644(4)
  & \cellcolor{mylightred} -541
  & \cellcolor{mylightred} -630(5)
  & \cellcolor{mylightred} -541\\
\hline

8 & -1e3
  & \cellcolor{mylightred} -709(13)
  & \cellcolor{mylightred} -543
  & \cellcolor{mylightred} -716(9)
  & \cellcolor{mylightred} -543\\
\hline

9 & 1e3
  & -540.33(8)
  & -540.16
  & -540.26(6)
  & -540.00\\
\hline

10 & -1e5
  & -6.52(3)e4
  & -6.15e4
  & -6.51(3)e4
  & -6.15e4\\
\hline

11 & 1e6
  & -1.347(4)e6
  & -1.237e6
  & -1.352(3)e6
  & -1.237e6\\
\hline

12 & -500
  & \cellcolor{mylightred} -789(9)
  & \cellcolor{mylightred} -541
  & \cellcolor{mylightred} -782(7)
  & \cellcolor{mylightred} -540\\
\hline

\makecell{13 \\ [-6pt] \scriptsize{$s=10$}} & 1e3
  & -395.18(5)
  & -394.11
  & -395.21(4)
  & -394.11\\
\hline

\makecell{14 \\ [-6pt] \scriptsize{$s=5$} \\ [-6pt] \scriptsize{$m=20$}} & 2e3
  & -558.108(4)
  & -557.567
  & -558.277(3)
  & -557.713\\
\hline

\makecell{15 \\ [-6pt] \scriptsize{$m=50$}} & 1e10
  & \cellcolor{mylightgreen} -2.1823(0)e9
  & \cellcolor{mylightgreen} -2.1823e9
  & \cellcolor{mylightgreen} -2.1823(0)e9
  & \cellcolor{mylightgreen} -2.1823e9\\
\hline

\makecell{16 \\ [-6pt] \scriptsize{$s=7$}  \\ [-6pt] \scriptsize{$m=-100$}} & 2e4
  & -1.2143(5)e4
  & -1.2088e4
  & -1.2140(2)e4
  & -1.2088e4\\
\hline

\end{tabular}}
\end{center}
\caption{Orders of magnitude of the tail cdf $p$ and pdf $f$ of the distributions in table \ref{tab:2_comp} (numbered in column 1), with extra parameters added to some of them, computed at far upper or lower tail points using the ray and tail methods. Green values are exact matches within uncertainty, red values are prominent disagreements.}
\label{tab:3_comp_tail}
\end{table}

Imhof \cite{imhof1961computing} provides a table that lists some upper tail cdf values of some generalized chi-square distributions with mixed-sign weights. In table \ref{tab:2_comp}, nos. 1-12, we list those `exact' values. No information is provided about their accuracy, so we assume that they are rounded to four places, i.e. their uncertainty is $\pm0.5$ in the last digit. Liu et al \cite{liu2009new} also provide a table that lists some upper tail cdf values of some generalized chi-square distributions with only positive weights (i.e. these are all infinite tails), computed with Ruben's exact method to a greater precision of 6 digits (nos. 13-16 here). In all cases, the parameters $s=m=0$. We compare the outputs from our implementations of the exact methods, i.e. Imhof, ray-trace and IFFT, against these exact values, using roughly the minimum settings needed to reach this precision.

Up to 4 total degrees of freedom, the ray method can be used with adaptive grid integral (quadrature), to the error tolerance that we specify. Beyond this it performs slower Monte-Carlo integration (on the GPU here), and reports the standard error of the output, and we increase the number of Monte-Carlo rays (mentioned in the table) until we reach the required precision. The IFFT method does not report uncertainty values, but we increase its grid span and number of grid points $N$ (mentioned in the table) until its output converges to the desired precision. Its values are rounded to the last digit.

Digits in parentheses at the end signify uncertainty, eg $(1)$ means $\pm1$ in the last digit. We see that in almost all cases, all our values match each other and the exact values within uncertainty. (If two numbers are $p_1 \pm \Delta p_1$ and $p_2 \pm \Delta p_2$, they agree if $\lvert p_1 - p_2 \rvert \leq \Delta p_1 + \Delta p_2$.) In two cases, Imhof's exact published values deviate slightly from the uncertainty range of our values (including our implementation of Imhof's method itself), and we believe that our values are more correct, since we compute them to greater precision and they match each other.

Imhof's method is by far the fastest here for the same accuracy, and therefore the best. This is for two reasons. First, here we compute values over only three points, whereas ray-trace and IFFT use vector operations that are faster when simultaneously computing values over many points. Second, these values were not too far in the tails and only had up to 6 decimal places, but when we are far in the tails or need much more accuracy, the Imhof integral slows down or quits with a wrong value, unlike ray-trace.

In table \ref{tab:3_comp_tail}, we compare the orders of magnitude of the tail cdf and pdf values of the distributions in table \ref{tab:2_comp} (with some $s$ and $m$ values added to some of them), computed by the ray method and the tail approximation. The points at which these values are computed are so far in the tails that the values are smaller than \code{realmin}, and these are the only two methods that work there. The ray method computations are repeated 10 times to get a mean and SEM which is used as the uncertainty.

We see that there are only a few exact agreements between the methods, even though they are often pretty close. In some cases though, the ray method's estimate is noticeably smaller than that of the tail method (highlighted in red). These disagreements are larger than the Monte-Carlo variance in the ray method, so they are systematic. Such errors are sometimes reduced when many more rays are used. So this is possibly because the parts of the domain containing probability mass are sometimes not sufficiently sampled by the rays, especially in this far tail. 

We also see that the orders of magnitude of the cdf and pdf at a point are always nearly identical in the far tail. As we showed in sec. \ref{sec:inf-tail}, this is because the asymptotic pdf and cdf in the far tail have the same expression.

\subsection{With randomly sampled parameters}

Bodenham et al \cite{bodenham2016comparison} have suggested that testing the accuracy of methods on a few points across some hand-picked distributions, as in the previous section, is not thorough enough, and instead recommend to randomly sample a large set of generalized chi-square parameters, then compare the methods across an array of quantile points for each distribution in the set. In this section we show performance tests of our exact methods along these lines.

We first randomly sample a broad set of parameters. We draw the number $n$ of constituent non-central chi-square terms uniformly from 1 through 10. Then we draw each of the $n$ weights $w_i$ from a standard normal distribution, and each of the $n$ degrees of freedom $k_i$ uniformly from 1 through 10. Each non-centrality $\lambda_i$ is with equal probability either 0 (so that term is a central chi-square), or equal to $10^z$, where $z$ is a standard normal, so that the non-centrality can spread across several orders of magnitude. Similarly, $s$ is with equal probability either 0 (so that there is no linear term), or equal to $10^z$. $m$ is also drawn similarly to $s$, except that its sign is positive or negative with equal probability.

With this sampling, almost all distributions generated are non-elliptical, i.e. with both tails infinite. We keep 2000 of these non-elliptical distributions, then we separately generate 2000 elliptical distributions with a lower finite tail, by keeping $s$ at 0, and taking the absolute values of the $w_i$. This allows us to test our methods equally over both elliptical and non-elliptical distributions.

\begin{figure}[!ht]
\includegraphics[width=\columnwidth]{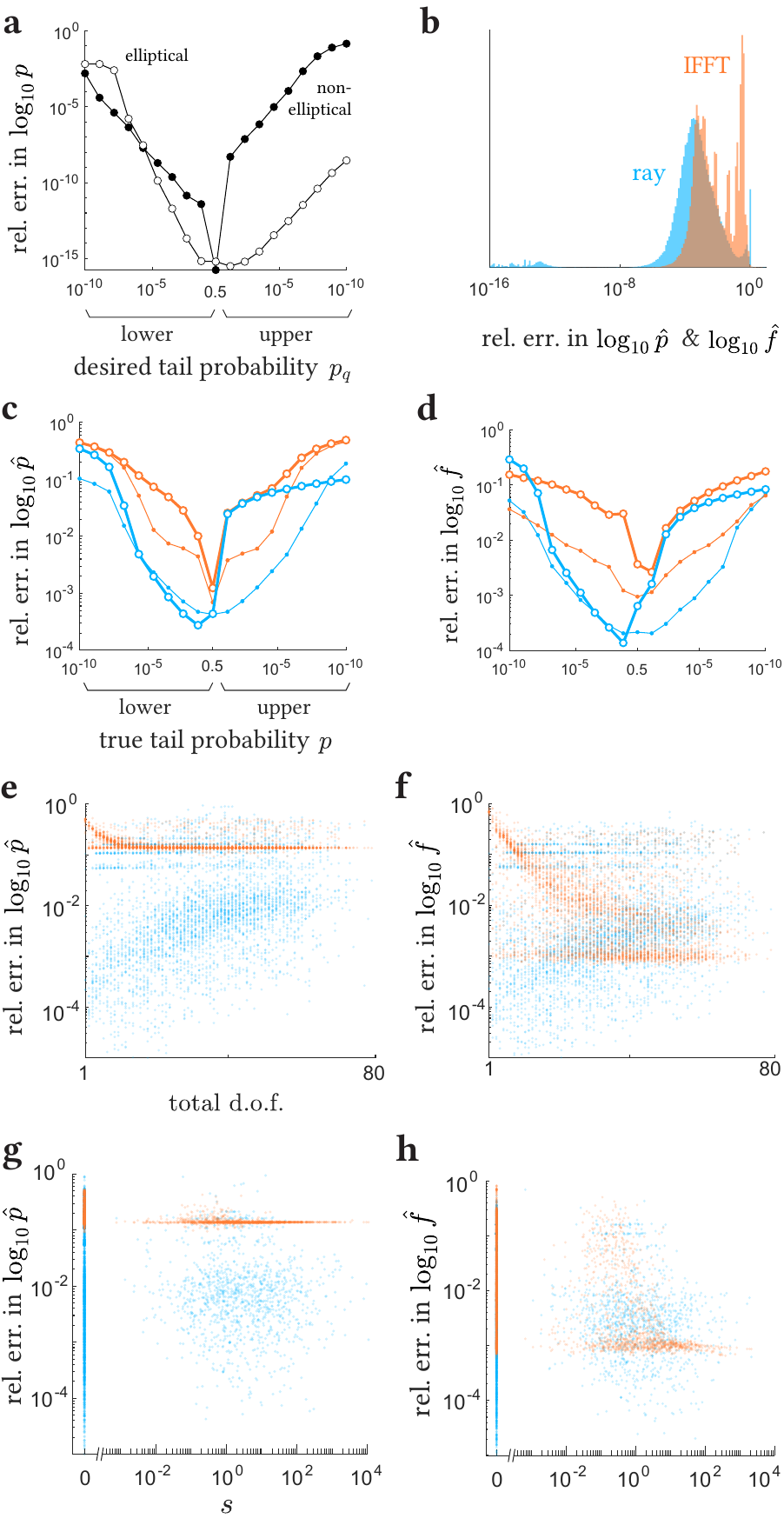}
    \caption{Computation errors in our two exact methods, ray and IFFT, across random distributions. \textbf{a:} Avg. relative discrepancy between the desired tail probability $p_q$ and the actual tail probability $p$ at each inverted quantile, for elliptical and non-elliptical distributions. \textbf{b:} Distribution of relative errors of the orders of the tail cdf $\hat{p}$ and pdf $\hat{f}$. \textbf{c-d:} Relative errors in estimating the tail cdf and the pdf, against an array of quantile points. Open and filled circles are for elliptical and non-elliptical distributions. \textbf{e-f:} Relative errors of cdf and pdf estimates against the total degrees of freedom of each distribution. \textbf{g-h:} Relative errors of cdf and pdf estimates against the value of $s$ of each distribution. Vertical stack on the left is for $s=0$.}
    \label{fig:7_random_test}
\end{figure}

Now we select an array of 19 quantile values $p_q$: these are 10 lower tail probability (cdf) values uniformly logarithmically spaced from $10^{-10}$ to 0.5 (the median), then upper tail probability (complementary cdf) values similarly arrayed from the median to $10^{-10}$. Now, for each of our 4000 distributions, we use our \code{gx2inv} function to find the quantile points corresponding to these probabilities $p_q$. \code{gx2inv} uses numerical root-finding to invert the \code{gx2cdf} function, whose cdf method is automatically selected based on the parameters, so for elliptical and non-elliptical distributions, it inverts Ruben's method and Imhof's method respectively. In fig. \ref{fig:7_random_test}a we first look at the accuracy of this inverse cdf function, by plotting the relative discrepancy between the desired quantile value $p_q$, and the true tail probability $p$ computed by Ruben's or Imhof's method at the inverted quantile point: $\lvert (\log_{10} p-\log_{10} p_q)  \big/ \log_{10} p_q) \rvert$. We measure the discrepancy in the order of the probability instead of the probability itself, because $p_q$ is arrayed across several scales, and we find that the discrepancies in the order of magnitude, and not the value itself, are what is more comparable across these different scales. If $p$ at the inverted quantile was wrongly computed as 0, we take the relative discrepancy to be 1.

Not surprisingly, we see that for the same settings, the inversion starts to become more erroneous in the tails. When inverting Imhof's method for non-elliptical distributions with both infinite tails, this rise in error is more symmetric, but when inverting Ruben's method for elliptical distributions, the error is much lower for the upper infinite tail than the lower finite tail.

At each of these quantile points, we now compute the tail probability and density with high accuracy, so that we have reliable ground-truth values: we use Ruben's method (with 1000 terms) and Imhof's method (with variable precision and a small relative error tolerance of $10^{-15}$) for the elliptical and non-elliptical distributions respectively. Ruben takes an avg. $\pm$ sd of $0.5 \pm 0.1$ s to compute all 19 cdf values of a distribution, and Imhof takes $8 \pm 5$ s. We then compute those cdf and pdf values using the ray and IFFT methods. The lower tail probabilities and densities here in the elliptical distributions are not small enough to use the ellipse approximation. We use ray with double precision, and $2{\times} 10^6$ rays for the cdf and $10^7$ rays for the pdf, and IFFT with a span of $10^5$ times the range from the lowest to the highest quantile, and $2{\times} 10^7$ grid points for the cdf, and a span of $4{\times}10^5$ times that range, and $10^8$ grid points for the pdf. These settings were selected roughly via trial and error on a few distributions, and are not sacrosanct, and with these settings ray and IFFT need roughly equal time. For computing all 19 cdf values for each distribution, ray took an avg. $\pm$ sd of $6 \pm 1$ s, and IFFT took $5 \pm 1$ s. For computing all the pdf values for each distribution, ray took $12 \pm 3$ s, and IFFT took $15 \pm 3$ s. 

We then compute the fractional error in estimating the order of magnitude of the probability and density: $\lvert (\log_{10} \hat{p}-\log_{10} p)  \big/ \log_{10} p) \rvert$ and $\lvert (\log_{10} \hat{f}-\log_{10} f)  \big/ \log_{10} f) \rvert$, where $p$ and $f$ are respectively the tail probability (cdf below median, and complementary cdf above median) and density computed by the ground-truth (Ruben or Imhof) method, and $\hat{p}$ and $\hat{f}$ are computed by ray and IFFT. If $\hat{p}$ or $\hat{f}$ was wrongly computed as 0, we again take the relative error to be 1. For a small fraction of cases, the ground-truth $p$ or $f$ was also wrongly computed as 0, and we omit these cases.

Across all 19 quantiles of the elliptical distributions with a finite tail, the average $\pm$ sd of the relative error of estimating the log tail probability was $0.03 \pm 0.49$ with the ray method, and $0.14 \pm 0.17$ by the IFFT method. On the non-elliptical distributions with two infinite tails, the relative estimation error was $0.07 \pm 0.23$ by the ray method, and $0.18 \pm 0.20$ by IFFT. So the ray method is better on average for the same computation time. The relative error of estimating the log probability density on the elliptical distributions was $0.05 \pm 0.19$ with the ray method, and $0.08 \pm 0.22$ with IFFT, and on the non-elliptical distributions it was $0.01 \pm 0.12$ with the ray method, and $0.01 \pm 0.07$ with IFFT. So for computing the density, ray and IFFT are on average equally accurate for the same time. Fig. \ref{fig:7_random_test}b shows the overall distributions of these relative estimation errors of the log probability and density across all the distributions. We see that the ray method is overall more accurate for the same computation time.

Now we look at the variation in the cdf and pdf accuracy across different parameters. Figs. \ref{fig:7_random_test}c and d show the relative estimation errors in cdf and pdf across the different quantile values. We see that overall the ray method is better. Both methods are best for computing cdf and pdf values in the middle of the distributions, and they become more erroneous for the same settings as we go out to smaller values in the tails, so to perform better there they will need higher settings (more rays for the ray method, and wider and denser grids for IFFT). The IFFT method is equally accurate on both tails for both elliptical and non-elliptical distributions. The ray method is also equally accurate on both infinite tails of the non-elliptical distributions, but its error rises more sharply on the finite tail of elliptical distributions, for reasons we have discussed before.

Fig. \ref{fig:7_random_test}e shows the average relative estimation error of the cdf across all quantiles of all the distributions, as a function of the total degrees of freedom of each distribution. We see that the ray method is overall better, but its error climbs with the total d.o.f. This makes sense since this is the number of dimensions of the space in which it is integrating, so a fixed number of rays would be sampling the space more sparsely as the dimension grows, and will miss more features of the integration domain. The error of the IFFT is higher, but more static across d.o.f, since it is always computing a one-dimensional integral of the characteristic function, regardless of the total d.o.f. Fig. \ref{fig:7_random_test}f shows similar results for computing the pdf, but now the IFFT method is relatively better, and in fact shows some improvement with increasing d.o.f.

Figs. \ref{fig:7_random_test}g and h show the performance variation as a function of the linear $s$ term. We see that in general, the value of $s$, whether 0 or across different scales, has little effect on the performance of the methods, although IFFT improves slightly with higher $s$, as particularly visible for the pdf. This could be because the factor of $e^{-s^2 t^2/2}$ in the characteristic function $\phi(t)$ suppresses its tails and shrinks its width, so it is captured better by an inverse Fourier transform with the same grid span and density.

$m$ is just an offset and we have checked that it has no bearing on performance of either of the methods, so we don't plot it here.

In summary, we see that across a wide sample of distributions, the ray method is overall better than IFFT for computing the cdf and pdf for the same speed, except in the finite tail.

\subsection{With equal-covariance multinormals}
We have introduced two new exact methods: IFFT and ray. We have shown that the IFFT method agrees with previous exact methods in the body of the distribution, and does not claim high accuracy far in the tails. The ray method extends far into the tails, where no other exact method can compute a ground-truth value in reasonable time. So here we show an accuracy test of the ray method in the far tail, using a special case where the ground-truth value is a manageable number, not too tiny or too large for machine representation, that can be exactly calculated: the discriminability index $d'$.

$d'$ is a measure of statistical separation between two distributions, and for two multinormals with equal covariance $\bm{\Sigma}$, it is given by the easily-calculable Mahalanobis distance between them: $d'=\sqrt{(\bm{\mu}_a-\bm{\mu}_b)'\bm{\Sigma}^{-1}(\bm{\mu}_a-\bm{\mu}_b)}$. In a previous paper \cite{das2021method,das2020method}, we had introduced the Bayes discriminability index $d'_b=-2Z\left(p_e\right)$ that can be calculated from the classification error rate $p_e$ of any two distributions in general, i.e. by computing the probability of each distribution on the wrong side of the classification boundary. For two equal-covariance multinormals, the classification boundary is a plane between them, and we can use the ray method to integrate the multinormals on either side of the boundary, and compute the error rate and the corresponding $d'_b$. In this equal-covariance case, $d'_b$ should equal the Mahalanobis distance $d'$, so we can use this to test the accuracy of the ray method as we separate the two multinormals far apart, which takes us into the tail. Even though the equal-covariance is a simple case, ray-tracing still uses its full algorithm here, so this is a test for the entire general method.

We take two trivariate normals with the following covariance matrix:
\begin{equation*}
\mathbf{\Sigma} = \begin{bmatrix}
1 & .5 & .7\\
.5 & 2 & 1\\
.7 & 1 & 3\\
\end{bmatrix}
\end{equation*}

\begin{figure}[!t]
\includegraphics[width=\columnwidth]{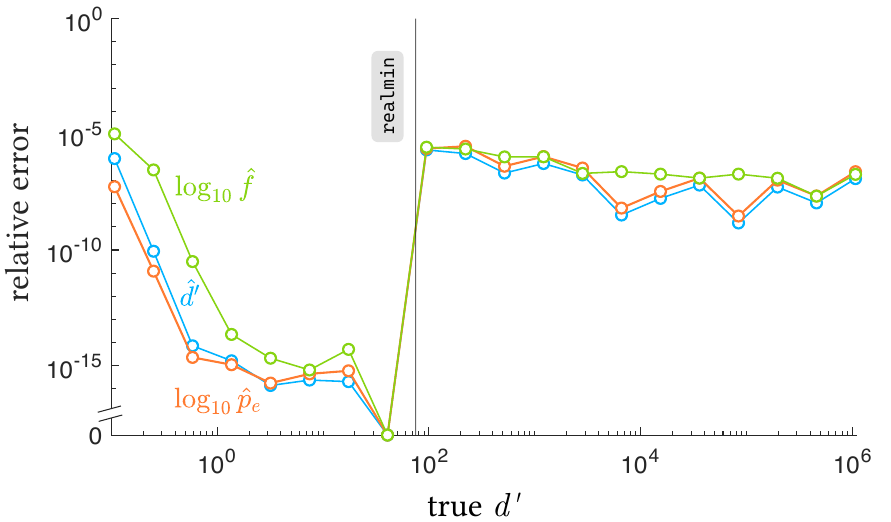}
    \caption{Relative errors in the ray method's estimates of the discriminability index $\hat{d}'$, and the orders of the classification error rate $\hat{p}_e$ between two multinormals, and the probability density $\hat{f}$ at the classification boundary, as we increase the true $d'$. Vertical line is the largest $d'$ computable in double precision (corresponding to $p_e=$ \code{realmin}), beyond which we use log calculations.}
    \label{fig:8_dprime_test}
\end{figure}

The first one stays at the origin while we move the mean of the second in logarithmic steps from the origin along the $\begin{bmatrix} 1 & 1 & 1 \end{bmatrix}$ vector. At each step we compute the true $d'$ using the Mahalanobis distance, and use the ray method to compute the classification error rate $\hat{p}_e$ between them, and the Bayes discriminability index $\hat{d}'$ from it. Fig. \ref{fig:8_dprime_test} shows the relative inaccuracy in this estimate $\frac{\lvert \hat{d}'-d' \rvert}{d'}$ with increasing distance. We first use the ray method with grid integration (with a relative tolerance of $10^{-20}$) to compute $\hat{p}_e$ down to the double-precision limit \code{realmin}, which corresponds to $\hat{d}' \approx 75$, already a rarely high value. For small $d'$ (<1), the estimate is slower to converge to an accurate value, but for large values, which are more relevant when measuring the performance of good classifiers such as ideal observers, the ray method is quite accurate with an error of only around $10^{-15}$.

In our previous paper \cite{das2021method,das2020method}, fig. 8a, we had reported this accuracy up to the double-precision limit of $d' \approx 75$, but now we can extend past it using our asymptotic log approximation. We use this with Monte-Carlo integration on $10^7$ rays to directly compute $\log_{10} \hat{p}_e$. From this, we can write an asymptotic expression for $\hat{d}'$, avoiding underflow. Suppose $\hat{d}'=2z$. Then, at large $z$,
\begin{equation*}
    \hat{p}_e =\bar{\Phi}(z) = \frac{e^{-z^2/2}}{z \sqrt{2 \pi}} \implies z^2 + 2 \ln z = \underbrace{-2 \ln (\hat{p}_e \sqrt{2 \pi})}_{\text{call this }w}.
\end{equation*}

So to 1st order, $z^2=w \implies 2 \ln z = \ln w$. Reinserting this gives us the 2nd order approximation : $\hat{d}'=2z = 2\sqrt{w-\ln w}$, where $w=-2 \ln 10 \ \log_{10} \hat{p}_e - \ln 2\pi$. Using this we can calculate $\hat{d}'$ up to a scale of $10^{154}$. We see in fig. \ref{fig:8_dprime_test} that even as the true $d'$ climbs very high, our estimate $\hat{d}'$ has relative errors of only around $10^{-8}$.

We can also directly test the accuracy in the ray method's estimate $\hat{p}_e$ of the classification error rate itself, which is a tail cdf, by comparing it to the true known error rate $p_e=\bar{\Phi}(d'/2)$. We noticed here that in the far tail, where the true $p_e$ was around $10^{-145643038}$ at a point, the estimated $\hat{p}_e$ was around $10^{-145821748}$. So in absolute terms, $\hat{p}_e$ was only $10^{-178710}$ of the true $p_e$ and egregiously wrong, and the estimation error would be 100\%. But arguably when comparing such tiny numbers, it is fairer to compare their orders of magnitude instead, which is off by only about 0.1\% here. We also notice that similar to our random-sample tests, it is the relative error in the log of the probability or density that is more comparable across scales. So in fig. \ref{fig:8_dprime_test} we show the relative error in the estimate of the order of magnitude of $p_e$, i.e. $\lvert (\log_{10} \hat{p}_e-\log_{10} p_e)  \big/ \log_{10} p_e) \rvert$, and we see that it closely follows the small estimation errors of the $d'$, which makes sense since $d'$ and $p_e$ are monotonic functions of each other.

We can also test the accuracy of computing the pdf by the ray method here, by computing the probability density of either of the multinormals at the planar classification boundary, which is the same as the density of a standard multinormal at a distance $d'/2$. The ray method is designed to compute the pdf of a \textit{function} of a multinormal though, not the multinormal itself, but we can work around that easily. We take the trivariate standard normal in $(x_1,x_2,x_3)$, and at each step of $d'$, we use the ray method to compute the pdf of the `function' $g(\bm{x})=x_1$ at $g=d'/2$ (again using grid integration with a relative tolerance of $10^{-20}$ in double precision, and Monte-Carlo integration with $10^7$ rays with the \code{log} precision option). We can compare this density $\hat{f}$ to the known true value of the density at the boundary, $f=\phi(d'/2)$. Like the cdf, this pdf is also tiny in the far tail, so we again look at the estimation error of its order of magnitude: $\lvert (\log_{10} \hat{f}-\log_{10} f)  \big/ \log_{10} f) \rvert$. We see that the accuracy is comparable to those of $\hat{d}'$ and $\log_{10} \hat{p}_e$.

To compute the tail probability $\hat{p}_e$ and $\hat{d}'$, with grid integration it takes from 1s (large $d'$) to 42s (small $d'$) per point, whereas Monte-Carlo integration on the GPU with $10^7$ rays takes about 7s per point. Computing the probability density $\hat{f}$ with grid integration in double precision takes from 1.5s (large $d'$) to 15s (small $d'$) per point, whereas Monte-Carlo integration on the GPU with $10^7$ rays takes around 2s per point.

\section{Conclusion}
In this paper we derived the mapping from generalized chi-square parameters to the parameters of the corresponding quadratic function of a multinormal, and presented two exact (IFFT and ray-trace) methods to compute the cdf and pdf of the distribution, and two approximate methods (ellipse and tail) for the finite and infinite tails of the distribution. These methods are all accompanied by our open-source Matlab and python implementations. We compared the performance of these methods against our implementations of previous exact methods by Ruben and Imhof to compute the cdf and pdf for elliptical and non-elliptical distributions. We showed how well they all agree in the middle of the distributions, how far into the tails they can reach, and their accuracy and speed, and presented a table of the best methods to use in different cases. We tested that our two new exact methods match previously published cdf values, and that they are accurate when tested across an array of quantile points on a large set of randomly sampled distributions. We also showed that the high-accuracy ray method can be used to correctly measure large discriminability indices $d'$ between two multinormal distributions.

The following are some avenues for future work: i.) completing the derivation for the infinite-tail approximation when $k_*$ is odd, ii.) seeing how the infinite-tail approximation compares to more recent moment-matching approximations, e.g. by Liu et al \cite{liu2009new} and Zhang et al \cite{zhang2022fast}, iii.) when computing the pdf/cdf of a multivariate polynomial function of a normal vector, the function along any ray is a polynomial, so all its roots can be found more easily than using a general numerical root-finding algorithm, iv.) extending the IFFT method to use the GPU, v.) the Monte-Carlo integral of the ray-tracing could be upgraded to some adaptive method that more quickly finds the parts of the domain that contain mass, and this could remedy its underestimates that we saw in table \ref{tab:3_comp_tail}.

\section{Acknowledgements}
I thank Dr Luis Mendo Tomás (Universidad Politécnica de Madrid) and Dr Andrew D. Horchler (Astrobotic) for their generous help with my Stack Overflow questions on variable precision, my longtime friend Dr Stefan Eccles (Okinawa Institute of Science and Technology) for backing up my multivariable calculus, and Thomas Hettasch (Technische Universiteit Delft) for correcting an error. I am grateful to my advisor Dr Wilson Geisler (The University of Texas at Austin) for allowing me the time and freedom to work on my own project.
\bibliographystyle{unsrt}
\bibliography{references}

\newpage
\appendix
\renewcommand\thesection{S}
\setcounter{figure}{0} \renewcommand\thefigure{S\arabic{figure}}
\setcounter{equation}{0} \renewcommand\theequation{S\arabic{equation}}

\end{document}